\documentclass{llncs}

\usepackage{chngpage}

\usepackage{booktabs}

\usepackage[utf8]{inputenc}

\usepackage{scrextend}

\usepackage[usenames, dvipsnames]{color}

\usepackage{tikz}
\usepackage{url}
\usepackage{amsmath, amssymb}
\usepackage{mathabx}
\usepackage{caption} 
\newcommand{\ms}[1]{\texttt{#1}}

\usepackage{fancyvrb}
\DefineVerbatimEnvironment{ex}{Verbatim}{numbers=left,numbersep=2mm,frame=single,fontsize=\scriptsize}

\usepackage{xspace}

\usepackage{tikz}

\usepackage{pifont}

\usepackage{tabularx}

\usepackage[colorinlistoftodos]{todonotes}

\newenvironment{table-1cols}{
  \scriptsize
  \sffamily
  \vspace{0.3cm}
  \begin{tabular}{l}
  \hline
  \textbf{Requirements} \\
  \hline

}{
  \hline
  \end{tabular}
  \linebreak
}

\newenvironment{table-2cols}{
  \scriptsize
  \sffamily
  \vspace{0.3cm}
  \begin{tabular}{l|l}
  \hline
  \textbf{Requirements} & \textbf{Covering DSCLs} \\
  \hline

}{
  \hline
  \end{tabular}
  \linebreak
}

\newenvironment{DL}{
  \vspace{0cm}
  \begin{tabular}{r l}

}{
  \end{tabular}
}

\newenvironment{constraint-languages-complexity}{
  \begin{tabular}{l|c|c|c|c|c|c}
  \hline
  \textbf{Complexity Class} & \textbf{DSP} & \textbf{OWL2-DL} & \textbf{OWL2-QL} & \textbf{ReSh} & \textbf{ShEx} & \textbf{SPIN} \\
  \hline

}{
  \hline
  \end{tabular}
  \linebreak
}

\newenvironment{user-fiendliness}{
  \begin{tabular}{l|c|c|c|c|c}
  \hline
  \textbf{criterion} & \textbf{DSP} & \textbf{OWL2} & \textbf{ReSh} & \textbf{ShEx} & \textbf{SPIN} \\
  \hline

}{
  \hline
  \end{tabular}
  \linebreak
}

\begin{document}
\renewcommand{\arraystretch}{1.3}
\title{Constraints to Validate RDF Data Quality on}
\subtitle{Common Vocabularies in the Social, Behavioral, and Economic Sciences}

\titlerunning{XXXXX}  
%
\author{Thomas Hartmann\inst{1} \and Benjamin Zapilko\inst{1} \and Joachim Wackerow\inst{1} \and Kai Eckert\inst{2}}
\authorrunning{} 
%
\institute{GESIS – Leibniz Institute for the Social Sciences, Germany\\
\email{\{firstname.lastname\}@gesis.org},\\ 
\and
University of Mannheim, Germany \\
\email{kai@informatik.uni-mannheim.de} 
}

\maketitle              

\begin{abstract}
To ensure high quality of and trust in both metadata and data, their representation in RDF must satisfy certain criteria - specified in terms of RDF constraints.
From 2012 to 2015 together with other Linked Data community members and experts from the social, behavioral, and economic sciences (\emph{SBE}), we developed diverse vocabularies to represent \emph{SBE} metadata and rectangular data in RDF.

The \emph{DDI-RDF Discovery Vocabulary (DDI-RDF)} 
is designed to support the dissemination, management,
and reuse of unit-record data, i.e., data about individuals, households, and businesses, collected in form of responses to studies and archived for research purposes.
The \emph{RDF Data Cube Vocabulary (QB)} is a W3C recommendation for expressing \emph{data cubes}, i.e. multi-dimensional aggregate data and its metadata. 
\emph{Physical Data Description (PHDD)} is a vocabulary to model data in rectangular format, i.e., tabular data. 
The data could either be represented in records with character-separated values (\emph{CSV}) or fixed length. 
The \emph{Simple Knowledge Organization System (SKOS)} is a vocabulary to build knowledge organization systems such as thesauri, classification schemes, and taxonomies.
\emph{XKOS} is a SKOS extension to describe formal statistical classifications. 

In this paper, we describe RDF constraints to validate metadata on unit-record data (\emph{DDI-RDF}), aggregated data (\emph{QB}), thesauri (\emph{SKOS}), and statistical classifications (\emph{XKOS}) and to validate tabular data (\emph{PHDD}) - all of them represented in RDF. We classified these constraints according to the severity of occurring constraint violations.
This technical report is updated continuously as modifying, adding, and deleting constraints remains ongoing work.

\end{abstract}

\section{Introduction}

For constraint formulation and RDF data validation, several languages exist or are currently developed. \emph{Shape Expressions (ShEx)}, \emph{Resource Shapes (ReSh)}, \emph{Description Set Profiles (DSP)}, \emph{OWL 2}, the \emph{SPARQL Inferencing Notation (SPIN)}, and \emph{SPARQL} are the six most promising and widely used constraint languages. OWL 2 is used as a constraint language under the closed-world and unique name assumptions. The W3C currently develops \emph{SHACL}, an RDF vocabulary for describing RDF graph structures. With its direct support of validation via SPARQL, SPIN is very popular and certainly plays an important role for future developments in this field. It is particularly interesting as a means to validate arbitrary constraint languages by mapping them to SPARQL \cite{BoschEckert2014-2}. Yet, there is no clear favorite and none of the languages is able to meet all requirements raised by data practitioners. Further research and development therefore is needed.

In 2013, the W3C organized the RDF Validation Workshop,\footnote{\url{http://www.w3.org/2012/12/rdf-val/}} 
where experts from industry, government, and academia discussed first use cases for constraint formulation and RDF data validation.
In 2014, two working groups on RDF validation have been established to develop a language to express constraints on RDF data: 
the \emph{W3C RDF Data Shapes Working Group}\footnote{\url{http://www.w3.org/2014/rds/charter}} (33 participants of 19 organizations) and the \emph{DCMI RDF Application Profiles Task Group}\footnote{\url{http://wiki.dublincore.org/index.php/RDF-Application-Profiles}} (29 people of 22 organizations) which among others bundles the requirements of data institutions of the cultural heritage sector and the \emph{social, behavioral, and economic (SBE)} sciences and represents them in the W3C group. 

Within the DCMI task group, a collaboratively curated database of RDF validation requirements\footnote{Online available at: \url{http://purl.org/net/rdf-validation}} has been created which contains the findings of the working groups based on various case studies provided by data institutions \cite{BoschEckert2014}. It is publicly available and open for further contributions.
The database connects requirements to use cases, case studies, and implementations and forms the basis of this paper. 
We distinguish 81 requirements to formulate constraints on RDF data; 
each of them corresponding to a constraint type.

In this paper, we collected constraints for commonly used vocabularies in the SBE domain (see Section \ref{sbe-vocabularies}), either from the vocabularies themselves or from domain and data experts, in order to gain a better understanding about the role of certain requirements for data quality and to direct the further development of constraint languages. 
We let the experts classify the constraints according to the severity of their violation. 

\section{Common Vocabularies in SBE Sciences}
\label{sbe-vocabularies}

We took all well-established and newly developed SBE vocabularies into account and defined constraints for three vocabularies commonly used in the SBE sciences which are briefly introduced in the following. We analyzed actual data according to constraint violations, as for these vocabularies large data sets are already published.

SBE sciences require high-quality data for their empirical research. For more than a decade, members of the SBE community have been developing and using a
metadata standard, composed of almost twelve hundred metadata fields, known as the \emph{Data Documentation Initiative (DDI)}, \footnote{\url{http://www.ddialliance.org/Specification/}}
an XML format to disseminate, manage,
and reuse data collected and archived for research \cite{Vardigan-2008}. 
In XML, the definition of schemas containing constraints and the validation of data according to these constraints is commonly used to ensure a certain level of data quality.
With the rise of the Web of Data, data professionals and institutions are very interested in having their data be discovered and used by publishing their data directly in RDF or at least publish accurate metadata about their data to facilitate data integration. Therefore, not only established vocabularies like SKOS are used; 
recently, members of the SBE and Linked Data community developed with the \emph{DDI-RDF Discovery Vocabulary (DDI-RDF)}\footnote{\url{http://rdf-vocabulary.ddialliance.org/discovery.html}} a means to expose \emph{DDI} metadata as Linked Data. 

The data most often used in research within SBE sciences is \emph{unit-record data}, i.e., data collected about individuals, businesses, and households, in form of responses to studies or taken from administrative registers such as hospital records, registers of births and deaths. A \emph{study} represents the process by which a data set was generated or collected. The range of unit-record data is very broad - including census, education, health data and business, social, and labor force surveys. This type of research data is held within data archives or data libraries after it has been collected, so that it may be reused by future researchers. By its nature, unit-record data is highly confidential and access is often only permitted for qualified researchers who must apply for access. Researchers typically represent their results as aggregated data in form of multi-dimensional tables with only a few columns: so-called \emph{variables} such as \emph{sex} or \emph{age}. Aggregated data, which answers particular research questions, is derived from unit-record data by statistics on groups or aggregates such as frequencies and arithmetic means. The purpose of publicly available aggregated data is to get a first overview and to gain an interest in further analyses on the underlying unit-record data. For more detailed analyses, researchers refer to unit-record data including additional variables needed to answer subsequent research questions. 

\emph{Formal childcare} is an example of an aggregated variable which captures the measured availability of childcare services in percent over the population in European Union member states by 
the dimensions \emph{year}, \emph{duration}, \emph{age} of the child, and \emph{country}.
Variables are constructed out of values (of one or multiple datatypes) and/or code lists.
The variable \emph{age}, e.g., may be represented by values of the datatype \emph{xsd:nonNegativeInteger} or by a code list of age clusters (e.g., '0 to 10' and '11 to 20'). 
The \emph{RDF Data Cube Vocabulary (QB)}\footnote{http://www.w3.org/TR/vocab-data-cube/} is a W3C recommendation for representing \emph{data cubes}, i.e., multi-dimensional aggregated data, in RDF \cite{Cyganiak2010}. 
A \emph{qb:DataStructureDefinition} contains metadata of the data collection.
The variable \emph{formal childcare} is modeled as \emph{qb:measure}, since it stands for what has been measured in the data collection.
\emph{Year}, \emph{duration}, \emph{age}, and \emph{country} are \emph{qb:dimensions}.
Data values, i.e., the availability of childcare services in percent over the population, are collected in a \emph{qb:DataSet}. 
Each data value is represented inside a \emph{qb:Observation} which contains values for each dimension. 

For more detailed analyses we refer to the underlying unit-record data. The aggregated variable \emph{formal childcare} is calculated on the basis of six unit-record variables (i.a., \emph{Education at pre-school}) for which detailed metadata is given (i.a., code lists) enabling researchers to replicate the results shown in aggregated data tables.
\emph{DDI-RDF} is used to represent metadata on unit-record data in RDF.
The study (\emph{disco:Study}) for which the unit-record data has been collected 
contains eight data sets (\emph{disco:LogicalDataSet})
including variables (\emph{disco:Variable}) like the six ones needed to calculate the variable \emph{formal childcare}.

The \emph{Simple Knowledge Organization System (SKOS)} is reused to a large extend to build SBE vocabularies.
The codes of the variable \emph{Education at pre-school} are modeled as \emph{skos:Concepts} and 
a \emph{skos:OrderedCollection} organizes them in a particular order within a \emph{skos:memberList}.
A variable may be associated with a theoretical concept (\emph{skos:Concept}) and \emph{skos:narrower} builds the hierarchy of theoretical concepts within a \emph{skos:ConceptScheme} of a study.
The variable \emph{Education at pre-school} is assigned to the theoretical concept \emph{Child Care} which is a narrower concept of the top concept \emph{Education}.
Controlled vocabularies (\emph{skos:ConceptScheme}), serving as extension and reuse mechanism,
organize types (\emph{skos:Concept}) of descriptive statistics (\emph{disco:SummaryStatistics}) like minimum, maximum, and arithmetic mean.

\section{Classification of RDF Constraints according to the Severity of Constraint Violations}

A concrete constraint is instantiated from one of the 81 constraint types and is defined for a specific vocabulary.
It does not make sense to determine the severity of constraint violations of an entire constraint type,
as the severity depends on the individual context and vocabulary.
SBE experts determined the default \emph{severity level}\footnote{The possibility to define severity levels in vocabularies is in itself a requirement (\emph{R-158}).} for each constraint to indicate how serious the violation of the constraint is. We use the classification system of log messages in software development like \emph{Apache Log4j 2} \cite{Apache-2015}, the \emph{Java Logging API},\footnote{\url{http://docs.oracle.com/javase/7/docs/api/java/util/logging/Level.html}} and the \emph{Apache Commons Logging API}\footnote{\url{http://commons.apache.org/proper/commons-logging/}} as many data practitioners also have experience in software development and software developers intuitively understand these levels. We simplify this commonly accepted classification system and distinguish the three severity levels (1) \emph{informational}, (2) \emph{warning}, and (3) \emph{error}.
Violations of \emph{informational} constraints point to desirable but not necessary data improvements to achieve RDF representations which are ideal in terms of syntax and semantics of used vocabularies. 
\emph{Warnings} are syntactic or semantic problems which typically should not lead to an abortion of data processing.
\emph{Errors}, in contrast, are syntactic or semantic errors which should cause the abortion of data processing. 
Although we provide default severity levels for each constraint, validation environments should enable users to adapt the severity levels of constraints according to their individual needs.

\section{RDF Constraints by RDF Constraint Type}

\subsection{Subsumption}

A \emph{subclass axiom}\footnote{{\em R-100-SUBSUMPTION}} ({\em concept inclusion} in DL) states that the class \emph{C1} is a subclass of the class \emph{C2} - \emph{C1} is more specific than \emph{C2}, 
i.e. each resource of the class \emph{C1} must also be part of the class extension of \emph{C2}.

\begin{itemize}
	\item \textbf{{\em DISCO-C-SUBSUMPTION-01:}} 
All {\em disco:Universe}s must also be {\em skos:Concept}s (\ms{Universe $\sqsubseteq$ Concept}).
	\begin{itemize}
		\item severity level: ERROR
	\end{itemize}
\end{itemize}

\subsection{Class Equivalence}

{\em Class Equivalence}\footnote{{\em R-3-EQUIVALENT-CLASSES}} asserts that two concepts have the same instances.
While synonyms are an obvious example of equivalent concepts, in practice one more
often uses concept equivalence to give a name to complex expressions \cite{Kroetzsch2012}.
Concept equivalence is indeed subsumption from left and right ($A \sqsubseteq B$ and $B \sqsubseteq A$ implies $A \equiv B$).

\begin{itemize}
	\item \textbf{{\em DISCO-C-CLASS-EQUIVALENCE-01:}}
All {\em sio:SIO\_000367} resources must also be {\em disco:Variable}s (\ms{Variable $\equiv$ SIO\_000367}).
The Semanticscience Integrated Ontology (SIO)\footnote{https://code.google.com/p/semanticscience/wiki/SIO} provides a simple, integrated ontology of types and relations for rich description of objects, processes and their attributes.
{\em sio:SIO\_000367} is a variable defined as a value that may change within the scope of a given problem or set of operations.
Thus, {\em sio:SIO\_000367} is equivalent to {\em disco:Variable}.
\begin{itemize}
	\item severity level: INFO
\end{itemize}
\end{itemize}

\subsection{Sub Properties}

{\em Sub Properties}\footnote{\emph{R-54-SUB-OBJECT-PROPERTIES}, \emph{R-64-SUB-DATA-PROPERTIES}} state that the property \emph{P1} is a sub property of the property \emph{P2} - that is, if an individual \emph{x} is connected by \emph{P1} to an individual or a literal \emph{y}, then \emph{x} is also connected by \emph{P2} to \emph{y}. 

\begin{itemize}
	\item \textbf{{\em DISCO-C-SUB-PROPERTIES-01:}}
If an individual \emph{x} is connected by {\em disco:fundedBy} to an individual \emph{y}, then \emph{x} is also connected by {\em dcterms:contributor} to \emph{y} (\ms{fundedBy $\sqsubseteq$ contributor}). 
	\begin{itemize}
		\item severity level: ERROR
	\end{itemize}
\end{itemize}

\subsection{Property Domains}

{\em Property Domains}\footnote{{\em R-25-OBJECT-PROPERTY-DOMAIN}, {\em R-26-DATA-PROPERTY-DOMAIN}} ({\em domain restrictions on roles} in DL) restrict the domain of object and data properties.
The purpose is to declare that a given property is associated with a class. 
In OO terms this is the declaration of a member, field, attribute or association. 
$\exists R. \top \sqsubseteq C$ is the object property restriction where $R$ is the object property (role) whose domain is restricted to concept $C$.

\begin{itemize}
	\item \textbf{{\em DISCO-C-PROPERTY-DOMAIN-01:}} 
	{\em Property Domain} constraints are defined for each \emph{Disco} object and data property.
  Only {\em disco:Question}s, e.g., can have {\em disco:responseDomain} relationships (\ms{$\exists$ responseDomain.$\top$ $\sqsubseteq$ Question}).
	\begin{itemize}
		\item Severity level: ERROR
	\end{itemize}
\end{itemize}

\begin{itemize}
	\item \textbf{{\em DATA-CUBE-C-PROPERTY-DOMAIN-01:}} 
	{\em Property Domain} constraints are defined for each \emph{Data Cube} object and data property.
  Only {\em qb:Observation}s, e.g., can have {\em qb:dataSet} relationships (\ms{$\exists$ dataSet.$\top$ $\sqsubseteq$ Observation}).
	\begin{itemize}
		\item Severity level: ERROR
	\end{itemize}
\end{itemize}

\begin{itemize}
	\item \textbf{{\em DCAT-C-PROPERTY-DOMAIN-01:}} 
	{\em Property Domain} constraints are defined for each \emph{DCAT} object and data property.
  Only {\em dcat:Catalog}s, e.g., can have {\em dcat:dataset} relationships (\ms{$\exists$ dataset.$\top$ $\sqsubseteq$ Catalog}).
	\begin{itemize}
		\item Severity level: ERROR
	\end{itemize}
\end{itemize}

\begin{itemize}
	\item \textbf{{\em PHDD-C-PROPERTY-DOMAIN-01:}} 
	{\em Property Domain} constraints are defined for each \emph{PHDD} object and data property.
  Only {\em phdd:Table}s, e.g., can have {\em phdd:isStructuredBy} relationships (\ms{$\exists$ isStructuredBy.$\top$ $\sqsubseteq$ Table}).
	\begin{itemize}
		\item Severity level: ERROR
	\end{itemize}
\end{itemize}

\begin{itemize}
	\item \textbf{{\em SKOS-C-PROPERTY-DOMAIN-01:}} 
	{\em Property Domain} constraints are defined for each \emph{SKOS} object and data property.
	Only {\em skos:ConceptSchemes}, e.g., can have {\em skos:hasTopConcept} relationships (\ms{$\exists$ hasTopConcept.$\top$ $\sqsubseteq$ ConceptScheme}).
	\begin{itemize}
		\item Severity level: ERROR
	\end{itemize}
\end{itemize}

\begin{itemize}
	\item \textbf{{\em XKOS-C-PROPERTY-DOMAIN-01:}} 
	{\em Property Domain} constraints are defined for each \emph{XKOS} object and data property.
	\begin{itemize}
		\item Severity level: ERROR
	\end{itemize}
\end{itemize}

\subsection{Property Ranges}

{\em Property Ranges}\footnote{{\em R-28-OBJECT-PROPERTY-RANGE}, {\em R-35-DATA-PROPERTY-RANGE}} ({\em range restrictions on roles} in DL) restrict the range of object and data properties.
$\top \sqsubseteq \forall R . C$ is the range restriction to the object property $R$ (restricted by the concept $C$). 

\begin{itemize}
	\item \textbf{{\em DISCO-C-PROPERTY-RANGES-01:}} 
	{\em Property Range} constraints are defined for each \emph{Disco} object and data property.
  {\em disco:caseQuantity} relationships, e.g., can only point to literals of the datatype {\em xsd:nonNegativeInteger} (\ms{$\top$ $\sqsubseteq$ $\forall$ caseQuantity.nonNegativeInteger}).
	\begin{itemize}
		\item Severity level: ERROR
	\end{itemize}
\end{itemize}

\begin{itemize}
	\item \textbf{{\em DATA-CUBE-C-PROPERTY-RANGES-01:}} 
	{\em Property Range} constraints are defined for each \emph{Data Cube} object and data property.
  {\em qb:order} relationships, e.g., can only point to literals of the datatype {\em xsd:string} (\ms{$\top$ $\sqsubseteq$ $\forall$ order.string}).
	\begin{itemize}
		\item Severity level: ERROR
	\end{itemize}
\end{itemize}

\begin{itemize}
	\item \textbf{{\em DCAT-C-PROPERTY-RANGES-01:}} 
	{\em Property Range} constraints are defined for each \emph{DCAT} object and data property.
  {\em dcat:bytes} relationships, e.g., can only point to literals of the datatype {\em xsd:integer} (\ms{$\top$ $\sqsubseteq$ $\forall$ bytes.integer}).
	\begin{itemize}
		\item Severity level: ERROR
	\end{itemize}
\end{itemize}

\begin{itemize}
	\item \textbf{{\em PHDD-C-PROPERTY-RANGES-01:}} 
	{\em Property Range} constraints are defined for each \emph{PHDD} object and data property.
  {\em phdd:caseQuantity} relationships, e.g., can only point to literals of the datatype {\em xsd:nonNegativeInteger} (\ms{$\top$ $\sqsubseteq$ $\forall$ caseQuantity.nonNegativeInteger}).
	\begin{itemize}
		\item Severity level: ERROR
	\end{itemize}
\end{itemize}

\begin{itemize}
	\item \textbf{{\em SKOS-C-PROPERTY-RANGES-01:}} 
	{\em Property Range} constraints are defined for each \emph{SKOS} object and data property.
	\begin{itemize}
		\item Severity level: ERROR
	\end{itemize}
\end{itemize}

\begin{itemize}
	\item \textbf{{\em XKOS-C-PROPERTY-RANGES-01:}} 
	{\em Property Range} constraints are defined for each \emph{XKOS} object and data property.
  {\em xkos:belongsTo} relationships, e.g., can only point to instances of the class {\em skos:Concept} (\ms{$\top$ $\sqsubseteq$ $\forall$ belongsTo.Concept}).
	\begin{itemize}
		\item Severity level: ERROR
	\end{itemize}
\end{itemize}

\subsection{Inverse Object Properties}

In many cases, properties are used bi-directionally and then accessed in the inverse direction, e.g. parent $\equiv$ child$^{-}$. There should be a way to declare value type, cardinality etc of those inverse relations without having to declare a new property URI. 
The object property \emph{OP1} is an inverse\footnote{\emph{R-56-INVERSE-OBJECT-PROPERTIES}} of the object property \emph{OP2}. 
Thus, if an individual \emph{x} is connected by \emph{OP1} to an individual \emph{y}, then \emph{y} is also connected by \emph{OP2} to \emph{x}, and vice versa.

\begin{itemize}
	\item \textbf{{\em DISCO-C-INVERSE-OBJECT-PROPERTIES-01}}:
	\emph{disco:CategoryStatistics} resources are accessed from codes (\emph{skos:Concept}s) via \emph{disco:statisticsCategory}$^{-}$.
	\begin{itemize}
		\item severity level: ERROR
	\end{itemize}
	\item \textbf{{\em DISCO-C-INVERSE-OBJECT-PROPERTIES-02}}:
	\emph{disco:SummaryStatistics} resources are accessed from \emph{disco:Variable}s via \emph{disco:statisticsVariable}$^{-}$.
	\begin{itemize}
		\item severity level: ERROR
	\end{itemize}
	\item \textbf{{\em DISCO-C-INVERSE-OBJECT-PROPERTIES-03}}:
	\emph{disco:Variable}s are accessed from \emph{disco:Question}s via \emph{disco:question}$^{-}$.
	\begin{itemize}
		\item severity level: ERROR
	\end{itemize}
\end{itemize}

\subsection{Symmetric Object Properties}

A role is symmetric if it is equivalent to its own inverse \cite{Kroetzsch2012}.
An object property symmetry axiom\footnote{\emph{R-61-SYMMETRIC-OBJECT-PROPERTIES}} states that the object property expression \emph{OPE} is symmetric - that is, if an individual \emph{x} is connected by \emph{OPE} to an individual \emph{y}, then \emph{y} is also connected by \emph{OPE} to \emph{x}. 	

\subsection{Asymmetric Object Properties}

A property is asymmetric\footnote{{\em R-62-ASYMMETRIC-OBJECT-PROPERTIES}} if it is disjoint from its own inverse \cite{Kroetzsch2012}.
An object property asymmetry axiom states that the object property \emph{OP} is asymmetric - that is, if an individual \emph{x} is connected by \emph{OP} to an individual \emph{y}, then \emph{y} cannot be connected by \emph{OP} to \emph{x}. 

\begin{itemize}
	\item \textbf{{\em DISCO-C-ASYMMETRIC-OBJECT-PROPERTIES-01:}} 
A {\em disco:Variable} may be based on a {\em disco:RepresentedVariable}.
A {\em disco:RepresentedVariable}, however, cannot be based on a {\em disco:Variable}.
This is a kind of mistake which may occur as a semantically equivalent object property for the other direction may also be possible ({\em disco:basisOf}) (\ms{$basedOn \sqcap basedOn^{-} \sqsubseteq \bot$}).
	\begin{itemize}
		\item severity level: ERROR
	\end{itemize}
\end{itemize}

\subsection{Reflexive Object Properties}

\emph{Reflexive Object Properties}\footnote{\emph{R-59-REFLEXIVE-OBJECT-PROPERTIES}} (\emph{reflexive roles}, \emph{global reflexivity} in DL) can be expressed by imposing local reflexivity on the top concept \cite{Kroetzsch2012}.

\subsection{Irreflexive Object Properties}

An object property is irreflexive\footnote{\emph{R-60-IRREFLEXIVE-OBJECT-PROPERTIES}} (\emph{irreflexive role} in DL) if it is never locally reflexive \cite{Kroetzsch2012}.
An object property irreflexivity axiom \emph{IrreflexiveObjectProperty( OPE )} states that the object property expression \emph{OPE} is irreflexive - that is, no individual is connected by \emph{OPE} to itself. 

\begin{itemize}
  \item \textbf{{\em DISCO-C-IRREFLEXIVE-OBJECT-PROPERTIES-01:}}
	In \emph{Disco}, every object property is irreflexive.
  No individual is connected by the object property {\em instrument} to itself (\ms{$\top$ $\sqsubseteq$ $\neg$ $\exists  instrument . Self$}).
	\begin{itemize}
		\item severity level: ERROR
	\end{itemize}
\end{itemize}

\subsection{Class-Specific Irreflexive Object Properties}

A property is \emph{irreflexive} if it is never locally reflexive \cite{Kroetzsch2012}.
An object property irreflexivity axiom states that the object property \emph{OP} is irreflexive - that is, no individual is connected by \emph{OP} to itself.
\emph{Class-Specific Irreflexive Object Properties} are object properties which are irreflexive within a given context, e.g. a class. 

\begin{itemize}
  \item \textbf{{\em DISCO-C-CLASS-SPECIFIC-IRREFLEXIVE-OBJECT-PROPERTIES-01:}}
Within the Disco context, {\em skos:Concept}s cannot be related via the object property {\em skos:boader} to themselves (\ms{Concept $\sqsubseteq$ $\neg$$\exists$ broader.Self.}). 
	\begin{itemize}
		\item severity level: ERROR
	\end{itemize}
	\item \textbf{{\em DISCO-C-CLASS-SPECIFIC-IRREFLEXIVE-OBJECT-PROPERTIES-02:}}
Within the Disco context, {\em skos:Concept}s cannot be related via the object property {\em skos:narrower} to themselves (\ms{Concept $\sqsubseteq$ $\neg$$\exists$ narrower.Self.}). 
	\begin{itemize}
		\item severity level: ERROR
	\end{itemize}
\end{itemize}

\subsection{Disjoint Properties}

A \emph{disjoint properties axiom}\footnote{\emph{R-9-DISJOINT-PROPERTIES}} states that all of the properties are pairwise disjoint; 
that is, no individual \emph{x} can be connected to an individual/literal \emph{y} by these properties. 

\begin{itemize}
	\item \textbf{{\em DATA-CUBE-C-DISJOINT-PROPERTIES-01:}} 
	All \emph{Data Cube} properties (not having the same domain and range classes) are defined to be pairwise disjoint.
  The properties \emph{qb:dataSet} and \emph{qb:structure} are disjoint ($dataSet \sqsubseteq \neg structure$).
	\begin{itemize}
		\item severity level: ERROR
	\end{itemize}
\end{itemize}

\begin{itemize}
	\item \textbf{{\em DCAT-C-DISJOINT-PROPERTIES-01:}} 
	All \emph{DCAT} properties (not having the same domain and range classes) are defined to be pairwise disjoint.
	\begin{itemize}
		\item severity level: ERROR
	\end{itemize}
\end{itemize}

\begin{itemize}
	\item \textbf{{\em DISCO-C-DISJOINT-PROPERTIES-01:}} 
	All \emph{Disco} properties (not having the same domain and range classes) are defined to be pairwise disjoint.
  The properties \emph{disco:variable} and \emph{disco:question} are disjoint ($variable \sqsubseteq \neg question$).
	\begin{itemize}
		\item severity level: ERROR
	\end{itemize}
\end{itemize}

\begin{itemize}
	\item \textbf{{\em PHDD-C-DISJOINT-PROPERTIES-01:}} 
	All \emph{PHDD} properties (not having the same domain and range classes) are defined to be pairwise disjoint.
	The properties \emph{phdd:isStructuredBy} and \emph{phdd:column} are disjoint ($isStructuredBy \sqsubseteq \neg column$)
	\begin{itemize}
		\item severity level: ERROR
	\end{itemize}
\end{itemize}

\begin{itemize}

	\item \textbf{{\em SKOS-C-DISJOINT-PROPERTIES-01:}} 
	All \emph{SKOS} properties (not having the same domain and range classes) are defined to be pairwise disjoint.
	\begin{itemize}
		\item severity level: ERROR
	\end{itemize}
	\item \textbf{{\em SKOS-C-DISJOINT-PROPERTIES-02\footnote{Corresponds to qSKOS Quality Issues - SKOS Semi-Formal Consistency Issues - Disjoint Labels Violation}:}}
	Disjoint Labels Violation:
  Covers condition S13 from the SKOS reference document stating that "\emph{skos:prefLabel}, \emph{skos:altLabel} and \emph{skos:hiddenLabel} are pairwise disjoint properties". 
	\begin{itemize}
	  \item Implementation:
		A SPARQL query collects all labels of all concepts, building an in-memory structure. This structure is then checked for disjoint entries. 
		\item severity level: ERROR
	\end{itemize}
	
\end{itemize}

\begin{itemize}
	\item \textbf{{\em XKOS-C-DISJOINT-PROPERTIES-01:}} 
	All \emph{XKOS} properties (not having the same domain and range classes) are defined to be pairwise disjoint.
	\begin{itemize}
		\item severity level: ERROR
	\end{itemize}
\end{itemize}

\subsection{Disjoint Classes}

{\em Disjoint Classes}\footnote{{\em R-7-DISJOINT-CLASSES}} state that all of the classes are pairwise disjoint; 
that is, no individual can be at the same time an instance of these disjoint classes.

\begin{itemize}
	\item \textbf{{\em DATA-CUBE-C-DISJOINT-CLASSES-01:}} 
All \emph{Data Cube} classes are defined to be pairwise disjoint.
\begin{itemize}
		\item severity level: ERROR
	\end{itemize}
\end{itemize}

\begin{itemize}
	\item \textbf{{\em DCAT-C-DISJOINT-CLASSES-01:}} 
All \emph{DCAT} classes are defined to be pairwise disjoint.
\begin{itemize}
		\item severity level: ERROR
	\end{itemize}
\end{itemize}

\begin{itemize}
	\item \textbf{{\em DISCO-C-DISJOINT-CLASSES-01:}} 
All \emph{Disco} classes are defined to be pairwise disjoint (e.g. \ms{Study $\sqcap$ Variable $\sqsubseteq$ $\perp$}).
\begin{itemize}
		\item severity level: ERROR
	\end{itemize}
\end{itemize}

\begin{itemize}
	\item \textbf{{\em PHDD-C-DISJOINT-CLASSES-01:}} 
All \emph{PHDD} classes are defined to be pairwise disjoint.
\begin{itemize}
		\item severity level: ERROR
	\end{itemize}
\end{itemize}

\begin{itemize}
	\item \textbf{{\em SKOS-C-DISJOINT-CLASSES-01:}} 
All \emph{SKOS} classes are defined to be pairwise disjoint.
\begin{itemize}
		\item severity level: ERROR
	\end{itemize}
\end{itemize}

\begin{itemize}
	\item \textbf{{\em XKOS-C-DISJOINT-CLASSES-01:}} 
All \emph{XKOS} classes are defined to be pairwise disjoint.
\begin{itemize}
		\item severity level: ERROR
	\end{itemize}
\end{itemize}

\subsection{Context-Specific Property Groups}

The \emph{Context-Specific Property Groups}\footnote{\emph{R-66-PROPERTY-GROUPS}} constraint groups data and object properties within a context (e.g. a class).

\subsection{Context-Specific Inclusive OR of Properties}

\emph{Inclusive or} is a logical connective joining two or more predicates that yields the logical value "true" when at least one of the predicates is true.
\emph{Context-Specific Inclusive OR of Properties}\footnote{\emph{R-202-CONTEXT-SPECIFIC-INCLUSIVE-OR-OF-PROPERTIES}} constraints specify that individuals are valid if they have at least one property relationship of one or multiple properties stated within a given context.
The context can be an application profile, a shape, or a class, i.e., the constraint applies for individuals of this specific class.

\subsection{Context-Specific Inclusive OR of Property Groups}

At least one property group must match for individuals of a specific context. 
Context may be a class, a shape, or an application profile.

\subsection{Recursive Queries}

Resource Shapes is a recursive language\footnote{\emph{R-222-RECURSIVE-QUERIES}} (the value shape of a Resource Shape is in turn another Resource Shape). 
There is no way to express that in SPARQL without hand-waving "and then you call the function again here" or "and then you embed this operation here" text.  
The embedding trick doesn't work
in the general case because SPARQL can't express recursive queries,
e.g. "test that this Issue is valid and all of the Issues that references, recursively".
Most SPARQL engines already have
functions that go beyond the official SPARQL 1.1 spec. 
The cost of that sounds manageable.

\subsection{Individual Inequality}

An \emph{individual inequality axiom}\footnote{\emph{R-14-DISJOINT-INDIVIDUALS}} DifferentIndividuals( a$_1$ ... a$_n$ ) states that all of the individuals a$_i$, 1 $\leq$ i $\leq$ n, are different from each other; 
that is, no individuals a$_i$ and a$_j$ with i $\neq$ j can be derived to be equal. 
This axiom can be used to axiomatize the unique name assumption — the assumption that all different individual names denote different individuals. 

\subsection{Equivalent Properties}

An \emph{equivalent object properties axiom}\footnote{\emph{R-4-EQUIVALENT-OBJECT-PROPERTIES}} \emph{EquivalentObjectProperties( OPE$_1$ ... OPE$_n$ )} states that all of the object property expressions \emph{OPE$_i$, 1 $\leq$ i $\leq$ n}, are semantically equivalent to each other. This axiom allows one to use each \emph{OPE$_i$} as a synonym for each \emph{OPE$_j$} — that is, in any expression in the ontology containing such an axiom, \emph{OPE$_i$} can be replaced with \emph{OPE$_j$} without affecting the meaning of the ontology. The axiom \emph{EquivalentObjectProperties( OPE$_1$ OPE$_2$ )} is equivalent to the following two axioms \emph{SubObjectPropertyOf( OPE$_1$ OPE$_2$ )} and \emph{SubObjectPropertyOf( OPE$_1$ OPE$_2$ )}.

An \emph{equivalent data properties axiom}\footnote{\emph{R-5-EQUIVALENT-DATA-PROPERTIES}} \emph{EquivalentDataProperties( DPE$_1$ ... DPE$_n$ )} states that all the data property expressions \emph{DPE$_i$, 1 $\leq$ i $\leq$ n}, are semantically equivalent to each other. This axiom allows one to use each \emph{DPE$_i$} as a synonym for each \emph{DPE$_j$} — that is, in any expression in the ontology containing such an axiom, DPE$_i$ can be replaced with \emph{DPE$_j$} without affecting the meaning of the ontology. The axiom \emph{EquivalentDataProperties( DPE$_1$ DPE$_2$ )} can be seen as a syntactic shortcut for the following axiom \emph{SubDataPropertyOf( DPE$_1$ DPE$_2$ )} and \emph{SubDataPropertyOf( DPE$_1$ DPE$_2$ )}.

\begin{itemize}
	\item \textbf{{\em DATA-CUBE-C-EQUIVALENT-PROPERTIES-01:}}
	Equivalent properties from different versions of \emph{Data Cube} can be marked as equivalent. 
	As a consequence, the properties can be replaced by each other without affecting the meaning.
\end{itemize}

\begin{itemize}
	\item \textbf{{\em DCAT-C-EQUIVALENT-PROPERTIES-01:}}
	Equivalent properties from different versions of \emph{DCAT} can be marked as equivalent. 
	As a consequence, the properties can be replaced by each other without affecting the meaning.
\end{itemize}

\begin{itemize}
	\item \textbf{{\em DISCO-C-EQUIVALENT-PROPERTIES-01:}}
	Equivalent properties from different versions of \emph{Disco} can be marked as equivalent, e.g. \emph{disco:containsVariable} and \emph{disco:variable}. 
	As a consequence, the properties can be replaced by each other without affecting the meaning.
\end{itemize}

\begin{itemize}
	\item \textbf{{\em PHDD-C-EQUIVALENT-PROPERTIES-01:}}
	Equivalent properties from different versions of \emph{PHDD} can be marked as equivalent. 
	As a consequence, the properties can be replaced by each other without affecting the meaning.
\end{itemize}

\begin{itemize}
	\item \textbf{{\em SKOS-C-EQUIVALENT-PROPERTIES-01:}}
	Equivalent properties from different versions of \emph{SKOS} can be marked as equivalent. 
	As a consequence, the properties can be replaced by each other without affecting the meaning.
\end{itemize}

\begin{itemize}
	\item \textbf{{\em XKOS-C-EQUIVALENT-PROPERTIES-01:}}
	Equivalent properties from different versions of \emph{XKOS} can be marked as equivalent. 
	As a consequence, the properties can be replaced by each other without affecting the meaning.
\end{itemize}

\subsection{Property Assertions}

\emph{{Property Assertions}}\footnote{\emph{R-96-PROPERTY-ASSERTIONS}}
and includes positive property assertions and negative property assertions.
A \emph{positive object property assertion ObjectPropertyAssertion( OPE a$_1$ a$_2$ )} states that the individual a$_1$ is connected by the object property expression \emph{OPE} to the individual a$_2$. 
A \emph{negative object property assertion NegativeObjectPropertyAssertion( OPE a$_1$ a$_2$ )} states that the individual a$_1$ is not connected by the object property expression \emph{OPE} to the individual a$_2$. 
A \emph{positive data property assertion DataPropertyAssertion( DPE a lt )} states that the individual a is connected by the data property expression DPE to the literal \emph{lt}. 
A \emph{negative data property assertion NegativeDataPropertyAssertion( DPE a lt )} states that the individual a is not connected by the data property expression \emph{DPE} to the literal \emph{lt}.

\subsection{Data Property Facets}

For datatype properties it should be possible to declare frequently needed \emph{facets}\footnote{\emph{R-46-CONSTRAINING-FACETS}} to drive user interfaces and validate input against simple conditions, including min/max value, regular expressions, string length - similar to XSD datatypes. 
Constraining facets, to restrict datatypes of RDF literals, may be: \emph{xsd:length}, \emph{xsd:minLength}, \emph{xsd:maxLength}, \emph{xsd:pattern}, \emph{xsd:enumeration}, \emph{xsd:whiteSpace}, \\ \emph{xsd:maxInclusive}, \emph{xsd:maxExclusive}, \emph{xsd:minExclusive}, \emph{xsd:minInclusive}, \emph{xsd:total} \emph{Digits}, \emph{xsd:fractionDigits}.

\begin{itemize}
	\item \textbf{{\em DISCO-C-DATA-PROPERTY-FACETS-01:}} The abstract of a series (\emph{dcterms:abstract}) should have a minimum length (\emph{xsd:minLength}) of some determined minimum length \emph{X}. 
	\begin{itemize}
		\item severity level: WARNING
	\end{itemize}
	\item \textbf{{\em DISCO-C-DATA-PROPERTY-FACETS-02:}} The abstract of a study (\emph{dcterms:abstract}) should have a minimum length (\emph{xsd:minLength}) of some determined minimum length \emph{X}. 
	\begin{itemize}
		\item severity level: WARNING
	\end{itemize}
\end{itemize}

\subsection{Literal Pattern Matching}

There are multiple use cases associated with the requirement to match literals according to given patterns\footnote{\emph{R-44-PATTERN-MATCHING-ON-RDF-LITERALS}}.

\begin{itemize}
	\item \textbf{{\em DISCO-C-LITERAL-PATTERN-MATCHING-01:}} Each \emph{disco:Variable} of a given \emph{disco:LogicalDataSet} must have a given prefix for its variable name (\emph{skos:notation}). 
	\begin{itemize}
		\item severity level: INFO
	\end{itemize}
\end{itemize}

\subsection{Negative Literal Pattern Matching}

Literals of given data properties within given contexts do not have to match given patterns\footnote{\emph{R-44-PATTERN-MATCHING-ON-RDF-LITERALS}}. 

\begin{itemize}
	\item \textbf{{\em DISCO-C-NEGATIVE-LITERAL-PATTERN-MATCHING-01:}} 
\end{itemize}

\subsection{Object Property Paths}

\emph{Object Property Paths}\footnote{\emph{R-55-OBJECT-PROPERTY-PATHS}} (or \emph{Object Property Chains} and in DL terminology \emph{complex role inclusion axiom} or \emph{role composition}) is the more complex form of sub properties. 
This axiom states that, if an individual \emph{x} is connected by a sequence of object property expressions \emph{OPE$_1$, ..., OPE$_n$} with an individual \emph{y}, then \emph{x} is also connected with \emph{y} by the object property expression \emph{OPE}.  
Role composition can only appear on the left-hand side of complex role inclusions \cite{Kroetzsch2012}.

\subsection{Intersection}

Concept inclusions allow us to state that all mothers are female and that
all mothers are parents, but what we really mean is that mothers are exactly the female
parents. DLs support such statements by allowing us to form complex concepts such as
the \emph{intersection}\footnote{\emph{R-15-CONJUNCTION-OF-CLASS-EXPRESSIONS}, \emph{R-16-CONJUNCTION-OF-DATA-RANGES}} (also called \emph{conjunction})
which denotes the set of individuals that are both female and parents. A complex concept
can be used in axioms in exactly the same way as an atomic concept, e.g., in the
equivalence Mother $\equiv$ Female $\sqcap$ Parent .

\subsection{Disjunction}

A \emph{union class expression}\footnote{{\em R-17-DISJUNCTION-OF-CLASS-EXPRESSIONS}, {\em R-18-DISJUNCTION-OF-DATA-RANGES}} contains all individuals that are instances of at least one class $C_{i}$ for 1 $\leq$ i $\leq$ n. 
A \emph{union data range} contains all tuples of literals that are contained in at least one data range $DR_{i}$ for 1 $\leq$ i $\leq$ n.
Synonyms of {\em disjunction} are {\em union} and {\em inclusive or}.

\begin{itemize}
	\item \textbf{{\em DISCO-C-DISJUNCTION-01:}} 
Only {\em disco:Variable}s or {\em disco:Question}s or {\em disco:RepresentedVariable}s can have {\em disco:concept} relationships to {\em skos:Concept}s.

\begin{DL}
Variable $\sqcup$ Question $\sqcup$ RepresentedVariable $\sqsubseteq$ $\forall$ concept.Concept \\
\end{DL}
\begin{itemize}
	\item severity level: ERROR
\end{itemize}
\end{itemize}

\subsection{Negation}

A \emph{complement class expression}\footnote{\emph{R-19-NEGATION-OF-CLASS-EXPRESSIONS}, \emph{R-20-NEGATION-OF-DATA-RANGES}} \emph{ObjectComplementOf( CE )} contains all individuals that are not instances of the class expression \emph{CE}. 

\subsection{Existential Quantifications}

An \emph{existential class expression}\footnote{{\em R-86-EXISTENTIAL-QUANTIFICATION-ON-PROPERTIES}} ({\em existential restriction} in DL terminology) contains all those individuals that are connected by the property \emph{P} to an individual \emph{x} that is an instance of the class \emph{C} or to literals that are in the data range \emph{DR}.

\begin{itemize}
	\item \textbf{{\em DISCO-C-EXISTENTIAL-QUANTIFICATIONS-01:}} 
There must be at least one {\em disco:universe} relationship from {\em disco:StudyGroups} to {\em disco:Universe} (\ms{StudyGroup $\sqsubseteq$ $\exists$ universe.Universe}).
	\begin{itemize}
		\item severity level: ERROR
	\end{itemize}
	\item \textbf{{\em DISCO-C-EXISTENTIAL-QUANTIFICATIONS-02:}} 
There must be at least one {\em disco:universe} relationship from {\em disco:Studies} to {\em disco:Universe} (\ms{Study $\sqsubseteq$ $\exists$ universe.Universe}).
	\begin{itemize}
		\item severity level: ERROR
	\end{itemize}
	\item \textbf{{\em DISCO-C-EXISTENTIAL-QUANTIFICATIONS-03:}} 
There may be a {\em disco:universe} relationship from {\em disco:RepresentedVariable} to {\em disco:Universe} (\ms{RepresentedVariable $\sqsubseteq$ $\exists$ universe.Universe}).
	\begin{itemize}
		\item severity level: INFO
	\end{itemize}
	\item \textbf{{\em DISCO-C-EXISTENTIAL-QUANTIFICATIONS-04:}} 
There may be a {\em disco:universe} relationship from {\em disco:Variable} to {\em disco:Universe} (\ms{Variable $\sqsubseteq$ $\exists$ universe.Universe}).
	\begin{itemize}
		\item severity level: INFO
	\end{itemize}
	\item \textbf{{\em DISCO-C-EXISTENTIAL-QUANTIFICATIONS-05:}} 
There may be a {\em disco:universe} relationship from {\em disco:Question} to {\em disco:Universe} (\ms{Question $\sqsubseteq$ $\exists$ universe.Universe}).
	\begin{itemize}
		\item severity level: INFO
	\end{itemize}
	\item \textbf{{\em DISCO-C-EXISTENTIAL-QUANTIFICATIONS-06:}} 
There may be a {\em disco:universe} relationship from {\em disco:LogicalDataSet} to {\em disco:Universe} (\ms{LogicalDataSet $\sqsubseteq$ $\exists$ universe.Universe}).
	\begin{itemize}
		\item severity level: INFO
	\end{itemize}
	\item \textbf{{\em DISCO-C-EXISTENTIAL-QUANTIFICATIONS-07:}} 
  There is no relationship (\emph{disco:ddifile}) to a DDI-XML file containing further information about the series (\emph{disco:StudyGroup}) for further analyses.
	\begin{itemize}
		\item severity level: INFO
	\end{itemize}
	\item \textbf{{\em DISCO-C-EXISTENTIAL-QUANTIFICATIONS-08:}} 
  There is no relationship (\emph{disco:ddifile}) to a DDI-XML file containing further information about the study (\emph{disco:Study}) for further analyses.
	\begin{itemize}
		\item severity level: INFO
	\end{itemize}
	\item \textbf{{\em DISCO-C-EXISTENTIAL-QUANTIFICATIONS-09:}} 
  It is important to know the kind of data (\emph{disco:kindOfData}) collected for a particular series (\emph{disco:StudyGroup}).
	Survey data, e.g., is much easier accessible than census data. 
	For census data, it is necessary to get in contact with the individual data archive and if data access is granted it may take months to actually get the data.
	\begin{itemize}
		\item severity level: INFO
	\end{itemize}
	\item \textbf{{\em DISCO-C-EXISTENTIAL-QUANTIFICATIONS-10:}}
  It is important to know the kind of data (\emph{disco:kindOfData}) collected for a particular study (\emph{disco:Study}).
	Survey data, e.g., is much easier accessible than census data. 
	For census data, it is necessary to get in contact with the individual data archive and if data access is granted it may take months to actually get the data.
	\begin{itemize}
		\item severity level: INFO
	\end{itemize}
	\item \textbf{{\em DISCO-C-EXISTENTIAL-QUANTIFICATIONS-11:}} 
  Information about the temporal coverage (\emph{dcterms:temporal}) of a series (\emph{disco:StudyGroup}) is of interest for particular queries 
	(e.g. to search for all series of a given year (temporal coverage) and for which data is collected in which countries (spatial coverage) about which topics (topical coverage)).
	\begin{itemize}
		\item severity level: INFO
	\end{itemize}
	\item \textbf{{\em DISCO-C-EXISTENTIAL-QUANTIFICATIONS-12:}} 
  Information about the spatial coverage (\emph{dcterms:spatial}) of a series (\emph{disco:StudyGroup}) is of interest for particular queries 
	(e.g. to search for all series of a given year (temporal coverage) and for which data is collected in which countries (spatial coverage) about which topics (topical coverage)).
	\begin{itemize}
		\item severity level: INFO
	\end{itemize}
	\item \textbf{{\em DISCO-C-EXISTENTIAL-QUANTIFICATIONS-13:}} 
  Information about the topical coverage (\emph{dcterms:subject}) of a series (\emph{disco:StudyGroup}) is of interest for particular queries 
	(e.g. to search for all series of a given year (temporal coverage) and for which data is collected in which countries (spatial coverage) about which topics (topical coverage)).
	\begin{itemize}
		\item severity level: INFO
	\end{itemize}
	\item \textbf{{\em DISCO-C-EXISTENTIAL-QUANTIFICATIONS-14:}} 
  Information about the temporal coverage (\emph{dcterms:temporal}) of a study (\emph{disco:Study}) is of interest for particular queries 
	(e.g. to search for all studies of a given year (temporal coverage) and for which data is collected in which countries (spatial coverage) about which topics (topical coverage)).
	\begin{itemize}
		\item severity level: INFO
	\end{itemize}
	\item \textbf{{\em DISCO-C-EXISTENTIAL-QUANTIFICATIONS-15:}} 
  Information about the spatial coverage (\emph{dcterms:spatial}) of a study (\emph{disco:Study}) is of interest for particular queries 
	(e.g. to search for all studies of a given year (temporal coverage) and for which data is collected in which countries (spatial coverage) about which topics (topical coverage)).
	\begin{itemize}
		\item severity level: INFO
	\end{itemize}
	\item \textbf{{\em DISCO-C-EXISTENTIAL-QUANTIFICATIONS-16:}} 
  Information about the topical coverage (\emph{dcterms:subject}) of a study (\emph{disco:Study}) is of interest for particular queries 
	(e.g. to search for all studies of a given year (temporal coverage) and for which data is collected in which countries (spatial coverage) about which topics (topical coverage)).
	\begin{itemize}
		\item severity level: INFO
	\end{itemize}
	\item \textbf{{\em DISCO-C-EXISTENTIAL-QUANTIFICATIONS-17:}} 
  Information about the temporal coverage (\emph{dcterms:temporal}) of a data set (\emph{disco:LogicalDataSet}) is of interest for particular queries 
	(e.g. to search for all data sets of a given year (temporal coverage) and for which data is collected in which countries (spatial coverage) about which topics (topical coverage)).
	\begin{itemize}
		\item severity level: INFO
	\end{itemize}
	\item \textbf{{\em DISCO-C-EXISTENTIAL-QUANTIFICATIONS-18:}} 
  Information about the spatial coverage (\emph{dcterms:spatial}) of a data set (\emph{disco:LogicalDataSet}) is of interest for particular queries 
	(e.g. to search for all data sets of a given year (temporal coverage) and for which data is collected in which countries (spatial coverage) about which topics (topical coverage)).
	\begin{itemize}
		\item severity level: INFO
	\end{itemize}
	\item \textbf{{\em DISCO-C-EXISTENTIAL-QUANTIFICATIONS-19:}} 
  Information about the topical coverage (\emph{dcterms:subject}) of a data set (\emph{disco:LogicalDataSet}) is of interest for particular queries 
	(e.g. to search for all data sets of a given year (temporal coverage) and for which data is collected in which countries (spatial coverage) about which topics (topical coverage)).
	\begin{itemize}
		\item severity level: INFO
	\end{itemize}
	\item \textbf{{\em DISCO-C-EXISTENTIAL-QUANTIFICATIONS-20:}} 
  Information about the temporal coverage (\emph{dcterms:temporal}) of a data file (\emph{disco:DataFile}) is of interest for particular queries 
	(e.g. to search for all data files of a given year (temporal coverage) and for which data is collected in which countries (spatial coverage) about which topics (topical coverage)).
	\begin{itemize}
		\item severity level: INFO
	\end{itemize}
	\item \textbf{{\em DISCO-C-EXISTENTIAL-QUANTIFICATIONS-21:}} 
  Information about the spatial coverage (\emph{dcterms:spatial}) of a data file (\emph{disco:DataFile}) is of interest for particular queries 
	(e.g. to search for all data files of a given year (temporal coverage) and for which data is collected in which countries (spatial coverage) about which topics (topical coverage)).
	\begin{itemize}
		\item severity level: INFO
	\end{itemize}
	\item \textbf{{\em DISCO-C-EXISTENTIAL-QUANTIFICATIONS-22:}} 
  Information about the topical coverage (\emph{dcterms:subject}) of a data file (\emph{disco:DataFile}) is of interest for particular queries 
	(e.g. to search for all data files of a given year (temporal coverage) and for which data is collected in which countries (spatial coverage) about which topics (topical coverage)).
	\begin{itemize}
		\item severity level: INFO
	\end{itemize}
	\item \textbf{{\em DISCO-C-EXISTENTIAL-QUANTIFICATIONS-23:}} 
  Information about creators (\emph{dcterms:creator}) (persons or organizations) of a series is important when searching for series of the same creators.
	\begin{itemize}
		\item severity level: INFO
	\end{itemize}
	\item \textbf{{\em DISCO-C-EXISTENTIAL-QUANTIFICATIONS-24:}} 
  Information about creators (\emph{dcterms:creator}) (persons or organizations) of a studies is important when searching for studies of the same creators.
	\begin{itemize}
		\item severity level: INFO
	\end{itemize}
	\item \textbf{{\em DISCO-C-EXISTENTIAL-QUANTIFICATIONS-25:}} 
  If summary statistics are collected for studies, detailed further analyses are possible. 
	\begin{itemize}
		\item severity level: INFO
	\end{itemize}
	\item \textbf{{\em DISCO-C-EXISTENTIAL-QUANTIFICATIONS-26:}} 
  If category statistics are collected for studies, detailed further analyses are possible. 
	\begin{itemize}
		\item severity level: INFO
	\end{itemize}
	\item \textbf{{\em DISCO-C-EXISTENTIAL-QUANTIFICATIONS-27:}} 
  If a study has no associated data sets, the actual description of the data is missing. 
	Eventually, it is very hard or even impossible to get access to the data.
	\begin{itemize}
		\item severity level: ERROR
	\end{itemize}
	\item \textbf{{\em DISCO-C-EXISTENTIAL-QUANTIFICATIONS-28:}} 
  If there is no data file for a given data set, the description of the data set and the containing study is not sufficient.
	\begin{itemize}
		\item severity level: WARNING
	\end{itemize}
	\item \textbf{{\em DISCO-C-EXISTENTIAL-QUANTIFICATIONS-29:}} 
	The case quantity measures how many cases are collected for a study.
  High case quantity (disco:caseQuantity), stated for data files, is an indicator for high statistical quality of the underlying study.
	It indicates how comprehensive the study is.
	\begin{itemize}
		\item severity level: WARNING
	\end{itemize}
	\item \textbf{{\em DISCO-C-EXISTENTIAL-QUANTIFICATIONS-30:}} 
  High variable quantity (disco:variableQuantity), stated for data files, is an indicator for high statistical quality of the underlying study.
	It indicates how comprehensive the study is.
	\begin{itemize}
		\item severity level: WARNING
	\end{itemize}
	\item \textbf{{\em DISCO-C-EXISTENTIAL-QUANTIFICATIONS-31:}} 
  High variable quantity (disco:variableQuantity), stated for data sets, is an indicator for high statistical quality of the underlying study.
	It indicates how comprehensive the study is.
	\begin{itemize}
		\item severity level: WARNING
	\end{itemize}
	\item \textbf{{\em DISCO-C-EXISTENTIAL-QUANTIFICATIONS-32:}} 
  There is no summary statistics type information (\emph{disco:summaryStatisticsType}) for a summary statistics resource.
	\begin{itemize}
		\item severity level: ERROR
	\end{itemize}
	\item \textbf{{\em DISCO-C-EXISTENTIAL-QUANTIFICATIONS-33:}} 
  There is no summary statistics value (\emph{rdf:value}) for a summary statistics resource.
	\begin{itemize}
		\item severity level: ERROR
	\end{itemize}
	\item \textbf{{\em DISCO-C-EXISTENTIAL-QUANTIFICATIONS-34:}} 
  There is no relationship to a variable (\emph{disco:statisticsVariable}) for a summary statistics resource.
	\begin{itemize}
		\item severity level: ERROR
	\end{itemize}
	\item \textbf{{\em DISCO-C-EXISTENTIAL-QUANTIFICATIONS-35:}} 
  Category statistics resources must be related (\emph{disco:statisticsCategory}) to codes/categories
	\begin{itemize}
		\item severity level: ERROR
	\end{itemize}
	\item \textbf{{\em DISCO-C-EXISTENTIAL-QUANTIFICATIONS-36:}} 
  Category statistics resources must have at minimum one value for either frequency, percentage, or cumulative percentage.
	\begin{itemize}
		\item severity level: ERROR
	\end{itemize}
	\item \textbf{{\em DISCO-C-EXISTENTIAL-QUANTIFICATIONS-37:}} 
  Codes should be associated with categories (human-readable labels).
	\begin{itemize}
		\item severity level: INFO
	\end{itemize}
	\item \textbf{{\em DISCO-C-EXISTENTIAL-QUANTIFICATIONS-38:}} 
  An instrument (\emph{disco:Instrument}) may have a link (\emph{disco:externalDocumentation}) to the questionnaire.
	\begin{itemize}
		\item severity level: INFO
	\end{itemize}
	\item \textbf{{\em DISCO-C-EXISTENTIAL-QUANTIFICATIONS-39:}} 
  Questions (\emph{disco:Question}) must have question texts (\emph{disco:questionText}).
	\begin{itemize}
		\item severity level: ERROR
	\end{itemize}
	\item \textbf{{\em DISCO-C-EXISTENTIAL-QUANTIFICATIONS-40:}} 
  Questions (\emph{disco:Question}) may have response domains (\emph{disco:responseDomain}).
	\begin{itemize}
		\item severity level: INFO
	\end{itemize}
	\item \textbf{{\em DISCO-C-EXISTENTIAL-QUANTIFICATIONS-41:}} 
  Questionnaires (\emph{disco:Questionnaire}) may contain (\emph{disco:question}) questions (\emph{disco:Question}).
	\begin{itemize}
		\item severity level: INFO
	\end{itemize}
	\item \textbf{{\em DISCO-C-EXISTENTIAL-QUANTIFICATIONS-42:}} 
  Questions (\emph{disco:Question}) may have question numbers (\emph{skos:prefLabel}).
	\begin{itemize}
		\item severity level: INFO
	\end{itemize}
	\item \textbf{{\em DISCO-C-EXISTENTIAL-QUANTIFICATIONS-43:}} 
  Variables (\emph{disco:Variable}) may have relationships (\emph{disco:question}) to questions (\emph{disco:Question}), as variables are created out of questions or calculated on the basis of other variables.
	\begin{itemize}
		\item severity level: INFO
	\end{itemize}
	\item \textbf{{\em DISCO-C-EXISTENTIAL-QUANTIFICATIONS-44:}} 
  Data sets (\emph{disco:LogicalDataSet}) may have (\emph{disco:variable}) variables (\emph{disco:Variable}).
	\begin{itemize}
		\item severity level: INFO
	\end{itemize}
	\item \textbf{{\em DISCO-C-EXISTENTIAL-QUANTIFICATIONS-45:}} 
  Variables (\emph{disco:Variable}) may have (\emph{disco:concept}) an associated theoretical concept (\emph{skos:Concept}).
	\begin{itemize}
		\item severity level: INFO
	\end{itemize}
	\item \textbf{{\em DISCO-C-EXISTENTIAL-QUANTIFICATIONS-46:}} 
  Each variable (\emph{disco:Variable}) should have (\emph{disco:representation}) a variable representation (\emph{disco:Representation}) which is either an ordered code list (\emph{skos:OrderedCollection}), an unordered code list (\emph{skos:ConceptScheme}) or a union of datatypes (\emph{rdfs:Datatype}).
	\begin{itemize}
		\item severity level: WARNING
	\end{itemize}
\end{itemize}

\begin{itemize}
	\item \textbf{{\em DATA-CUBE-C-EXISTENTIAL-QUANTIFICATIONS-01:}}
	Dimensions have range (\emph{IC-4} \cite{CyganiakReynolds2014}) - 
	Every dimension declared in a \emph{qb:DataStructureDefinition} must have a declared \emph{rdfs:range}. 
	\begin{itemize}
		\item severity level: ERROR
	\end{itemize}
	\item \textbf{{\em DATA-CUBE-C-EXISTENTIAL-QUANTIFICATIONS-02:}}
	Concept dimensions have code lists (\emph{IC-5} \cite{CyganiakReynolds2014}) - 
	Every dimension with range \emph{skos:Concept} must have a \emph{qb:codeList}. 
	\begin{itemize}
		\item severity level: ERROR
	\end{itemize}
  \item \textbf{{\em DATA-CUBE-C-EXISTENTIAL-QUANTIFICATIONS-03:}}
	DSD includes measure (\emph{IC-3} \cite{CyganiakReynolds2014}) -  
	Every \emph{qb:DataStructureDefinition} must include (\emph{qb:component}, \emph{qb:componentProperty}) at least one declared measure. 
	\begin{itemize}
		\item severity level: ERROR
	\end{itemize}
	\item \textbf{{\em DATA-CUBE-C-EXISTENTIAL-QUANTIFICATIONS-04	:}}
	Slice Keys must be declared (\emph{IC-7} \cite{CyganiakReynolds2014}) -
	Every \emph{qb:SliceKey} must be associated with (\emph{qb:sliceKey}) a \emph{qb:DataStructureDefinition} (\ms{SliceKey $\sqsubseteq$ $\exists$ sliceKey$^{-}$.DataStructureDefinition}). 
	\begin{itemize}
		\item severity level: ERROR
	\end{itemize}
\end{itemize}

\subsection{Universal Quantifications}

A \emph{universal class expression}\footnote{{\em R-91-UNIVERSAL-QUANTIFICATION-ON-PROPERTIES}} ({\em value restriction} in DL) contains all those individuals that are connected by an object property only to individuals that are instances of a particular class.

\begin{itemize}
	\item \textbf{{\em DATA-CUBE-C-UNIVERSAL-QUANTIFICATIONS-01:}}
	\emph{Universal quantifications} are defined for each \emph{Data Cube} object and data property.
	\begin{itemize}
		\item Severity level: ERROR
	\end{itemize}
\end{itemize}

\begin{itemize}
	\item \textbf{{\em DCAT-C-UNIVERSAL-QUANTIFICATIONS-01:}}
	\emph{Universal quantifications} are defined for each \emph{DCAT} object and data property.
	Only {\em dcat:Catalogs} can have {\em dcat:dataset} relationships to {\em dcat:Datasets} (\ms{Catalog $\sqsubseteq$ $\forall$ dataset.Dataset}).
	\begin{itemize}
		\item Severity level: ERROR
	\end{itemize}
\end{itemize}

\begin{itemize}
	\item \textbf{{\em DISCO-C-UNIVERSAL-QUANTIFICATIONS-01:}}
	\emph{Universal quantifications} are defined for each \emph{Disco} object and data property.
  Only {\em disco:LogicalDataSet}s can have {\em disco:aggregation} relationships to {\em qb:DataSet}s (\ms{LogicalDataSet $\sqsubseteq$ $\forall$ aggregation.DataSet}).
	\begin{itemize}
		\item Severity level: ERROR
	\end{itemize}
\end{itemize}

\begin{itemize}
	\item \textbf{{\em PHDD-C-UNIVERSAL-QUANTIFICATIONS-01:}}
	\emph{Universal quantifications} are defined for each \emph{PHDD} object and data property.
	\begin{itemize}
		\item Severity level: ERROR
	\end{itemize}
\end{itemize}

\begin{itemize}
	\item \textbf{{\em SKOS-C-UNIVERSAL-QUANTIFICATIONS-01:}}
	\emph{Universal quantifications} are defined for each \emph{SKOS} object and data property.
	\begin{itemize}
		\item Severity level: ERROR
	\end{itemize}
\end{itemize}

\begin{itemize}
	\item \textbf{{\em XKOS-C-UNIVERSAL-QUANTIFICATIONS-01:}}
	\emph{Universal quantifications} are defined for each \emph{XKOS} object and data property.
	\begin{itemize}
		\item Severity level: ERROR
	\end{itemize}
\end{itemize}

\subsection{Minimum Unqualified Cardinality Restrictions}

A \emph{minimum cardinality restriction}\footnote{{\em R-81-MINIMUM-UNQUALIFIED-CARDINALITY-ON-PROPERTIES}, {\em R-211-CARDINALITY-CONSTRAINTS}} contains all those individuals that are connected by a property to at least n different individuals/literals 
that are instances of a particular class or data range. If the class is missing, it is taken to be \emph{owl:Thing}. 
If the data range is missing, it is taken to be \emph{rdfs:Literal}.
$\leq n R. \top$ is the minimum unqualified cardinality restriction where $n \in \mathbb{N}$ (written $\leq  n R$ in short).
For unqualified cardinality restrictions, classes respective data ranges are not stated.

\subsection{Minimum Qualified Cardinality Restrictions}

A \emph{minimum cardinality restriction}\footnote{{\em R-75-MINIMUM-QUALIFIED-CARDINALITY-ON-PROPERTIES}, {\em R-211-CARDINALITY-CONSTRAINTS}} contains all those individuals that are connected by a property to at least n different individuals/literals 
that are instances of a particular class or data range. If the class is missing, it is taken to be \emph{owl:Thing}. 
If the data range is missing, it is taken to be \emph{rdfs:Literal}.
$\geq n R. C$ is a minimum qualified cardinality restriction where $n \in \mathbb{N}$.

\begin{itemize}
	\item \textbf{{\em DATA-CUBE-C-MINIMUM-QUALIFIED-CARDINALITY-RESTRICTIONS-01:}}
	{\em Minimum Qualified Cardinality Restrictions} constraints are defined for each \emph{Data Cube} object and data property.
	\begin{itemize}
		\item Severity level: ERROR
	\end{itemize}
		\item \textbf{{\em DATA-CUBE-C-MINIMUM-QUALIFIED-CARDINALITY-RESTRICTIONS-02:}}
	Unique data set (\emph{IC-1} \cite{CyganiakReynolds2014}) -  
	Every \emph{qb:Observation} has (\emph{qb:dataSet}) exactly one associated \emph{qb:DataSet} (\ms{Observation $\sqsubseteq$ $\geq$1 dataSet.DataSet $\sqcap$ $\leq$1 dataSet.DataSet}). 
	\begin{itemize}
		\item Severity level: ERROR
	\end{itemize}
\end{itemize}

\begin{itemize}
	\item \textbf{{\em DCAT-C-MINIMUM-QUALIFIED-CARDINALITY-RESTRICTIONS-01:}}
	{\em Minimum Qualified Cardinality Restrictions} constraints are defined for each \emph{DCAT} object and data property.
	\begin{itemize}
		\item Severity level: ERROR
	\end{itemize}
\end{itemize}

\begin{itemize}
	\item \textbf{{\em DISCO-C-MINIMUM-QUALIFIED-CARDINALITY-RESTRICTIONS-01:}}
	{\em Minimum Qualified Cardinality Restrictions} constraints are defined for each \emph{Disco} object and data property.
  A {\em disco:Questionnaire}, e.g., has at least one {\em disco:question} relationship to {\em disco:Questions} (\ms{Questionnaire $\sqsubseteq$ $\geq$1 question.Question}).
	\begin{itemize}
		\item Severity level: ERROR
	\end{itemize}
\end{itemize}

\begin{itemize}
	\item \textbf{{\em PHDD-C-MINIMUM-QUALIFIED-CARDINALITY-RESTRICTIONS-01:}}
	{\em Minimum Qualified Cardinality Restrictions} constraints are defined for each \emph{PHDD} object and data property.
	\begin{itemize}
		\item Severity level: ERROR
	\end{itemize}
\end{itemize}

\subsection{Maximum Unqualified Cardinality Restrictions}

A \emph{maximum cardinality restriction} contains all those individuals that are connected by a property to at most n different individuals/literals that are instances of a particular class or data range. If the class is missing, it is taken to be \emph{owl:Thing}. If the data range is not present, it is taken to be \emph{rdfs:Literal}.
Unqualified means that the class respective the data range is not stated. 
$\geq n R. \top$ is a \emph{maximum unqualified cardinality restriction}\footnote{\emph{R-82-MAXIMUM-UNQUALIFIED-CARDINALITY-ON-PROPERTIES}, \emph{R-211-CARDINALITY-CONSTRAINTS}} where $n \in \mathbb{N}$ (written $\geq  n R$ in short).

\subsection{Maximum Qualified Cardinality Restrictions}

A \emph{maximum cardinality restriction} contains all those individuals that are connected by a property to at most n different individuals/literals that are instances of a particular class or data range. If the class is missing, it is taken to be \emph{owl:Thing}. If the data range is not present, it is taken to be \emph{rdfs:Literal}.
Qualified means that the class respective the data range is stated. 
$\leq n R. C$ is a \emph{maximum qualified cardinality restriction}\footnote{\emph{R-76-MAXIMUM-QUALIFIED-CARDINALITY-ON-PROPERTIES}, \emph{R-211-CARDINALITY-CONSTRAINTS}} where $n \in \mathbb{N}$.

\begin{itemize}
	\item \textbf{{\em DISCO-C-MAXIMUM-QUALIFIED-CARDINALITY-RESTRICTIONS-01:}}
	A {\em disco:Variable} has at most one {\em disco:concept} relationship to a theoretical concept ({\em skos:Concept}) (\ms{Variable $\sqsubseteq$ $\leq$1 concept.Concept}).
	\begin{itemize}
		\item Severity level: ERROR
	\end{itemize}
	\item \textbf{{\em DATA-CUBE-C-MAXIMUM-QUALIFIED-CARDINALITY-RESTRICTIONS-01:}}
	Unique data set (\emph{IC-1} \cite{CyganiakReynolds2014}) -  
	Every \emph{qb:Observation} has (\emph{qb:dataSet}) exactly one associated \emph{qb:DataSet} (\ms{Observation $\sqsubseteq$ $\geq$1 dataSet.DataSet $\sqcap$ $\leq$1 dataSet.DataSet}). 
	\begin{itemize}
		\item Severity level: ERROR
	\end{itemize}
\end{itemize}

\subsection{Exact Unqualified Cardinality Restrictions}

An \emph{exact cardinality restriction}\footnote{{\em R-80-EXACT-UNQUALIFIED-CARDINALITY-ON-PROPERTIES}, {\em R-211-CARDINALITY-CONSTRAINTS}} contains all those individuals that are connected by a property to exactly n different individuals that are instances of a particular class or data range. 
If the class is missing, it is taken to be \emph{owl:Thing}. 
If the data range is not present, it is taken to be \emph{rdfs:Literal}.
Unqualified means that the class respective data range is not stated. 
$\geq n R. \top \sqcap \leq n R. \top $ is an exact unqualified cardinality restriction where $n \in \mathbb{N}$.

\begin{itemize}
	\item \textbf{{\em DATA-CUBE-C-EXACT-UNQUALIFIED-CARDINALITY-RESTRICTIONS-01:}}
	Unique slice structure (\emph{IC-9} \cite{CyganiakReynolds2014}) -  
	Each \emph{qb:Slice} must have exactly one associated \emph{qb:sliceStructure}. 
	\begin{itemize}
		\item Severity level: ERROR
	\end{itemize}
\end{itemize}

\subsection{Exact Qualified Cardinality Restrictions}

An \emph{exact cardinality restriction}\footnote{{\em R-74-EXACT-QUALIFIED-CARDINALITY-ON-PROPERTIES}, {\em R-211-CARDINALITY-CONSTRAINTS}} contains all those individuals that are connected by a property to exactly n different individuals that are instances of a particular class or data range. 
If the class is missing, it is taken to be \emph{owl:Thing}. 
If the data range is not present, it is taken to be \emph{rdfs:Literal}.
$\geq n R. C \sqcap \leq n R. C $ is an exact qualified cardinality restriction where $n \in \mathbb{N}$.

\begin{itemize}
	\item \textbf{{\em DISCO-C-EXACT-QUALIFIED-CARDINALITY-RESTRICTIONS-01:}}
A {\em disco:Question} has exactly 1 {\em disco:universe} relationship to {\em disco:Universe} (\ms{Question $\sqsubseteq$ $\geq$1 universe.Universe $\sqcap$ $\leq$1 universe.Universe}).
  \begin{itemize}
		\item Severity level: ERROR
	\end{itemize}
\end{itemize}

\begin{itemize}
	\item \textbf{{\em DATA-CUBE-C-EXACT-QUALIFIED-CARDINALITY-RESTRICTIONS-02:}}
	Unique DSD (\emph{IC-2} \cite{CyganiakReynolds2014}) -  
	Every \emph{qb:DataSet} has (\emph{qb:structure}) exactly one associated \emph{qb:DataStructureDefinition} (\ms{DataSet $\sqsubseteq$ $\geq$1 structure.DataStructureDefinition $\sqcap$ $\leq$1 structure.DataStructureDefinition}). 
	\begin{itemize}
		\item Severity level: ERROR
	\end{itemize}
\end{itemize}

\subsection{Transitive Object Properties}

\emph{Transitivity} is a special form of \emph{complex role inclusion}.
An \emph{object property transitivity axiom}\footnote{\emph{R-63-TRANSITIVE-OBJECT-PROPERTIES}} states that the object property is transitive — that is, if an individual \emph{x} is connected by the object property to an individual \emph{y} that is connected by the object property to an individual \emph{z}, then \emph{x} is also connected by the object property to \emph{z}.

\subsection{Context-Specific Exclusive OR of Properties}		

\emph{Exclusive or} is a logical operation that outputs true whenever both inputs differ (one is true, the other is false).
Only one of multiple properties within some context (e.g. a class, a shape, or an  application profile) leads to valid data\footnote{\emph{R-11-CONTEXT-SPECIFIC-EXCLUSIVE-OR-OF-PROPERTIES}}.
This constraint is generally expressed in DL as follows: \ms{$C \sqsubseteq (\neg A \sqcap B) \sqcup (A \sqcap \neg B)$}.

\subsection{Context-Specific Exclusive OR of Property Groups}

\emph{Exclusive or} is a logical operation that outputs true whenever both inputs differ (one is true, the other is false).
Only one of multiple property groups leads to valid data\footnote{{\em R-13-DISJOINT-GROUP-OF-PROPERTIES-CLASS-SPECIFIC}}.

\begin{itemize}
	\item \textbf{{\em DISCO-C-CONTEXT-SPECIFIC-EXCLUSIVE-OR-OF-PROPERTY-GROUPS-01:}}
Within the context of \emph{Disco}, \emph{skos:Concept}s can have either \emph{skos:definition} (when interpreted as theoretical concepts) or \emph{skos:notation} and \emph{skos:prefLabel} properties (when interpreted as codes and categories), but not both.
\begin{DL}
Concept $\sqsubseteq$ ($\neg$ D $\sqcap$ C) $\sqcup$ (D $\sqcap$ $\neg$ C) \\ 
D $\equiv$ A $\sqcap$ B \\
A $\sqsubseteq$ $\geq$ 1 notation.string $\sqcap$ $\leq$ 1 notation.string \\
B $\sqsubseteq$ $\geq$ 1 prefLabel.string $\sqcap$ $\leq$ 1 prefLabel.string \\
C $\sqsubseteq$ $\geq$ 1 definition.string $\sqcap$ $\leq$ 1 definition.string \\
\end{DL}
\begin{itemize}
		\item severity level: INFO
	\end{itemize}
\end{itemize}

\subsection{Allowed Values}

It is a common requirement to narrow down the value space of a property by an exhaustive enumeration of the valid values (both literals or resources). 
This is often rendered in drop down boxes or radio buttons in user interfaces. 
\emph{Allowed values}\footnote{{\em R-30-ALLOWED-VALUES-FOR-RDF-OBJECTS} and 
{\em R-37-ALLOWED-VALUES-FOR-RDF-LITERALS}} for properties can be IRIs, IRIs (matching one or multiple patterns), (any) literals, literals of a list of allowed literals (e.g. 'red' 'blue' 'green'), typed literals of one or multiple type(s) (e.g. \emph{xsd:string}).

\begin{itemize}
	\item \textbf{{\em DISCO-C-ALLOWED-VALUES-01}}.
{\em disco:CategoryStatistics} can only have {\em disco:computationBase} relationships to the values \emph{valid} and \emph{invalid} of the datatype {\em rdf:langString} (\ms{CategoryStatistics $\equiv$ $\forall$ computationBase.\{valid,invalid\} $\sqcap$ langString}).
	\begin{itemize}
		\item severity level: ERROR
	\end{itemize}
\end{itemize}

\subsection{Not Allowed Values}

A matching triple has any literal / object except those explicitly excluded\footnote{\emph{R-33-NEGATIVE-OBJECT-CONSTRAINTS}, \emph{R-200-NEGATIVE-LITERAL-CONSTRAINTS}}.

\subsection{Literal Ranges}

\emph{P1} is a data property (of an instance of class \emph{C1}) and its literal value must be between the range of [$V_{min}$,$V_{max}$]\footnote{{\em R-45-RANGES-OF-RDF-LITERAL-VALUES}}.

\begin{itemize}
	\item \textbf{{\em DISCO-C-LITERAL-RANGES-01:}}
{\em disco:percentage} (domain: {\em disco:CategoryStatistics}) literals must be of the datatype {\em xsd:double} whose range should be restricted to be between 0 and 100.
	\begin{itemize}
		\item severity level: ERROR
	\end{itemize}
	
	\item \textbf{{\em DISCO-C-LITERAL-RANGES-02:}}
{\em disco:cumulativePercentage} (domain: {\em disco:CategoryStatistics}) literals must be of the datatype {\em xsd:double} whose range should be restricted to be between 0 and 100.
	\begin{itemize}
		\item severity level: ERROR
	\end{itemize}

\end{itemize}

\subsection{Negative Literal Ranges}

\emph{P1} is a data property (of an instance of class \emph{C1}) and its literal value must not be between the range of [$V_{min}$,$V_{max}$]\footnote{{\em R-142-NEGATIVE-RANGES-OF-RDF-LITERAL-VALUES}}.

\subsection{Required Properties}

Properties may be required\footnote{\emph{R-68-REQUIRED-PROPERTIES}}.

\subsection{Optional Properties}

Properties may be optional\footnote{\emph{R-69-OPTIONAL-PROPERTIES}}.

\subsection{Repeatable Properties}

Properties may be repeatable\footnote{\emph{R-70-REPEATABLE-PROPERTIES}}.

\subsection{Negative Property Constraints}

Instances of a specific class must not have some object property\footnote{\emph{R-52-NEGATIVE-OBJECT-PROPERTY-CONSTRAINTS}, \emph{R-53-NEGATIVE-DATA-PROPERTY-CONSTRAINTS}}.

\subsection{Individual Equality}

\emph{Individual equality}\footnote{\emph{R-6-EQUIVALENT-INDIVIDUALS}} states that two different names are known to refer to the same individual \cite{Kroetzsch2012}.

\subsection{Functional Properties}

An \emph{object property functionality axiom}\footnote{\emph{R-57-FUNCTIONAL-OBJECT-PROPERTIES}} \emph{FunctionalObjectProperty( OPE )} states that the object property expression \emph{OPE} is functional — that is, for each individual \emph{x}, there can be at most one distinct individual \emph{y} such that \emph{x} is connected by \emph{OPE} to \emph{y}. Each such axiom can be seen as a syntactic shortcut for the following axiom: \emph{SubClassOf( owl:Thing ObjectMaxCardinality( 1 OPE ) )}.

\subsection{Inverse-Functional Properties}

An \emph{object property inverse functionality axiom}\footnote{\emph{R-58-INVERSE-FUNCTIONAL-OBJECT-PROPERTIES}} \emph{InverseFunctionalObjectProperty( OPE )} states that the object property expression \emph{OPE} is inverse-functional - that is, for each individual \emph{x}, there can be at most one individual \emph{y} such that \emph{y} is connected by \emph{OPE} with \emph{x}. Each such axiom can be seen as a syntactic shortcut for the following axiom: \emph{SubClassOf( owl:Thing ObjectMaxCardinality( 1 ObjectInverseOf( OPE ) ) )}.

\begin{itemize}
	\item \textbf{{\em DISCO-C-INVERSE-FUNCTIONAL-PROPERTIES-01:}}
	For each \emph{rdfs:Resource x}, there can be at most one distinct \emph{rdfs:Resource y} such that \emph{y} is connected by \emph{adms:identifier} to \emph{x} $(\ms{funct identifier}\sp{\overline{\ }})$.
	\begin{itemize}
		\item severity level: ERROR
	\end{itemize}
	\item \textbf{{\em DISCO-C-INVERSE-FUNCTIONAL-PROPERTIES-02:}}
	Keys are even more general than inverse-functional properties,
as a key can be a data, an object property, or a chain of properties \cite{Schneider2009}.
For this generalization purposes, as there are different sorts of key, and as keys can lead to undecidability, 
DL is extended with \emph{key boxes} and a special \emph{keyfor} construct (\ms{identifier \ms{keyfor} Resource}) \cite{Lutz2005}.
OWL 2 \emph{hasKey} implements \emph{keyfor} and thus can be used to identify resources uniquely, to merge resources with identical key property values, and to recognize constraint violations.
	\begin{itemize}
		\item severity level: ERROR
	\end{itemize}
\end{itemize}

\subsection{Value Restrictions}

\emph{Individual Value Restrictions}\footnote{\emph{R-88-VALUE-RESTRICTIONS}}: A has-value class expression \emph{ObjectHasValue( OPE a )} consists of an object property expression \emph{OPE} and an individual \emph{a}, and it contains all those individuals that are connected by \emph{OPE} to \emph{a}. Each such class expression can be seen as a syntactic shortcut for the class expression \emph{ObjectSomeValuesFrom( OPE ObjectOneOf( a ) )}. 
\emph{Literal Value Restrictions}: A has-value class expression \emph{DataHasValue( DPE lt )} consists of a data property expression \emph{DPE} and a literal \emph{lt}, and it contains all those individuals that are connected by \emph{DPE} to \emph{lt}. Each such class expression can be seen as a syntactic shortcut for the class expression \emph{DataSomeValuesFrom( DPE DataOneOf( lt ) )}.

\subsection{Self Restrictions}

A \emph{self-restriction} \emph{ObjectHasSelf( OPE )} consists of an object property expression \emph{OPE}, and it contains all those individuals that are connected by \emph{OPE} to themselves. 

\subsection{Primary Key Properties}

The \emph{Primary Key Properties}\footnote{R-226-PRIMARY-KEY-PROPERTIES} constraint is often useful to declare a given (datatype) property as the "primary key" of a class, so that a system can enforce uniqueness and also automatically build URIs from user input and data imported from relational databases or spreadsheets.. 
Starfleet officers, e.g., are uniquely identified by their command authorization code (e.g. to activate and cancel auto-destruct sequences).
It means that the property \emph{commandAuthorizationCode} is inverse functional - mapped to DL as follows:
$(\ms{funct commandAuthorizationCode}\sp{\overline{\ }})$
Keys, however, are even more general, i.e., a generalization of inverse functional properties \cite{Schneider2009}.
A key can be a datatype property, an object property, or a chain of properties.
For this generalization purposes, as there are different sorts of key, and as keys can lead to undecidability, 
DL is extended with \emph{key boxes} and a special \emph{keyfor} construct\cite{Lutz2005}.
This leads to the following DL mapping (only one simple property constraint):
\ms{commandAuthorizationCode \ms{keyfor} StarfleetOfficer}

\begin{itemize}
	\item see \emph{inverse-functional properties}
\end{itemize}

\subsection{Class-Specific Property Range}		

{\em Class-Specific Property Range}\footnote{{\em R-29-CLASS-SPECIFIC-RANGE-OF-RDF-OBJECTS}, {\em R-36-CLASS-SPECIFIC-RANGE-OF-RDF-LITERALS}} restricts the range of object and data properties for individuals within a specific context (e.g. class, shape, application profile).
The values of each member property of a class may be limited by their value type, such as \emph{xsd:string} or \emph{foaf:Person}. 

\begin{itemize}
	\item \textbf{{\em DISCO-C-CLASS-SPECIFIC-PROPERTY-RANGE-01:}}
Only {\em disco:Question}s can have {\em disco:questionText} relationships to literals of the datatype {\em rdf:langString} (\ms{$\neg$Question $\sqsubseteq$ $\neg\exists$ questionText.langString}).
	\begin{itemize}
		\item severity level: ERROR
	\end{itemize}
\end{itemize}

\subsection{Class-Specific Reflexive Object Properties}

Using DL terminology \emph{Class-Specific Reflexive Object Properties} is called local reflexivity - a set of individuals (of a specific class) that are related to themselves via a given role \cite{Kroetzsch2012}.

\subsection{Membership in Controlled Vocabularies}

Resources can only be members of listed controlled vocabularies\footnote{{\em R-32-MEMBERSHIP-OF-RDF-OBJECTS-IN-CONTROLLED-VOCABULARIES}, 
{\em R-39-MEMBERSHIP-OF-RDF-LITERALS-IN-CONTROLLED-VOCABULARIES}}.

\begin{itemize}
	\item \textbf{{\em DISCO-C-MEMBERSHIP-IN-CONTROLLED-VOCABULARIES-01}}:
{\em disco:SummaryStatistics} can only have {\em disco:summaryStatisticType} relationships to {\em skos:Concept}s which must be members of the controlled vocabulary {\em ddicv:SummaryStatisticType} which is a {\em skos:ConceptScheme}.

\begin{DL}
SummaryStatistics $\sqsubseteq$ $\forall summaryStatisticType.A$ \\
$A \equiv Concept \sqcap \forall inScheme . B$ \\
$B \equiv ConceptScheme \sqcap \{SummaryStatisticType\}$
\end{DL}

\begin{itemize}
		\item severity level: ERROR
	\end{itemize}
\end{itemize}

\begin{itemize}
	\item \textbf{{\em DATA-CUBE-C-MEMBERSHIP-IN-CONTROLLED-VOCABULARIES-01:}}
	Codes from code list (\emph{IC-19} \cite{CyganiakReynolds2014}) - 
	If a dimension property has a \emph{qb:codeList}, then the value of the dimension property on every \emph{qb:Observation} must be in the code list. 
	\begin{itemize}
		\item severity level: ERROR
	\end{itemize}
\end{itemize}

\subsection{IRI Pattern Matching}

IRI pattern matching applied on subjects, properties, and objects\footnote{\emph{R-21-IRI-PATTERN-MATCHING-ON-RDF-SUBJECTS}, \emph{R-22-IRI-PATTERN-MATCHING-ON-RDF-OBJECTS}, \emph{R-23-IRI-PATTERN-MATCHING-ON-RDF-PROPERTIES}}.

\begin{itemize}
	\item \textbf{{\em DISCO-C-IRI-PATTERN-MATCHING-01}}: \emph{disco:Study} resources must match a given IRI pattern.
	\begin{itemize}
		\item severity level: INFO
	\end{itemize}
\end{itemize}

\subsection{Literal Value Comparison}

Depending on the property semantics,
there are cases where two different literal values must have
a specific ordering with respect to an operator. 
\emph{P1} and \emph{P2} are the datatype properties we need to compare and 
\emph{OP} is the comparison operator (\textless, \textless=, \textgreater, \textgreater=, =, !=)\footnote{{\em R-43-LITERAL-VALUE-COMPARISON}}.
The {\em COMP Pattern}, one of the Data Quality Test Patterns, can be used to validate the {\em Literal Value Comparison} constraint \cite{Kontokostas2014}:

\begin{ex}
SELECT ?s WHERE { 
    ?s %%P1%% ?v1 .
    ?s %%P2%% ?v2 .
    FILTER ( ?v1 %%OP%% ?v2 ) }
\end{ex}

\begin{itemize}
	\item \textbf{{\em DISCO-C-LITERAL-VALUE-COMPARISON-01}}:
{\em disco:startDate}s must be before (‘\textless’) {\em disco:endDate}s.
To validate this constraint we bind the variables as follows (P1: {\em disco:startDate}, P2: {\em disco:endDate}, OP: \textless). 
	\begin{itemize}
		\item severity level: ERROR
	\end{itemize}
\end{itemize}

\subsection{Ordering}

With this constraint objects of object properties can be ordered as well as literals of data properties\footnote{\emph{R-121-SPECIFY-ORDER-OF-RDF-RESOURCES}, \emph{R-217-DEFINE-ORDER-FOR-FORMS/DISPLAY}}.

In DDI, variables, questions, and codes/categories are typically organized in a particular order. 
For obtaining this order, {\em skos:OrderedCollection} resources are used. 

\begin{itemize}
	\item \textbf{{\em DISCO-C-ORDERING-01}}: If \emph{disco:Variable}s of a given \emph{disco:LogicalDataSet} should be ordered, a collection of variables must be present in the data and connected with the data set. The collection of variables is of the type {\em skos:OrderedCollection} containing multiple variables (each represented as {\em skos:Concept}) in a {\em skos:memberList}. 
	\begin{itemize}
		\item severity level: INFO
	\end{itemize}
	\item \textbf{{\em DISCO-C-ORDERING-02}}: If \emph{disco:Question}s of a given \emph{disco:Questionnaire} should be ordered, a collection of questions must be present in the data and connected with the questionnaire. The collection of questions is of the type {\em skos:OrderedCollection} containing multiple questions (each represented as {\em skos:Concept}) in a {\em skos:memberList}. 
	\begin{itemize}
		\item severity level: INFO
	\end{itemize}
	\item \textbf{{\em DISCO-C-ORDERING-03}}: If codes/categories (\emph{skos:Concept}s) of a given \emph{disco:Representation} of a given \emph{disco:Variable} should be ordered, the variable representation should also be of the type {\em skos:OrderedCollection} containing multiple codes/categories (each represented as {\em skos:Concept}) in a {\em skos:memberList}. 
	\begin{itemize}
		\item severity level: INFO
	\end{itemize}
\end{itemize}

\subsection{Validation Levels}

Different levels of severity (priority)\footnote{\emph{R-205-VARYING-LEVELS-OF-ERROR}, \emph{R-135-CONSTRAINT-LEVELS}, \emph{R-158-SEVERITY-LEVELS-OF-CONSTRAINT-VIOLATIONS}, \emph{R-193-MULTIPLE-CONSTRAINT-VALIDATION-EXECUTION-LEVELS}} should be assigned to constraints.
Possible validation levels could be: informational, warning, error, fail, should, recommended, must, may, optional, closed (only this) constraints, open (at least this) constraint.

For \emph{Disco} each constraint should be assigned to exactly one \emph{validation level}.

\subsection{String Operations}

Many different \emph{string operations}\footnote{\emph{R-194-PROVIDE-STRING-FUNCTIONS-FOR-RDF-LITERALS}} are possible.
Some constraints require building new strings out of other strings.
Calculating the string length would also be another constraint of this type.

\begin{itemize}
	\item \textbf{{\em DISCO-C-STRING-OPERATIONS-01}}: The title of a study (\emph{dcterms:title}) (e.g. 'EU-SILC 2005') may be calculated out of the title of the containing series (\emph{dcterms:title}) (e.g. 'EU-SILC') and the human-readable label of the study (\emph{rdfs:label}) (e.g. '2005').   
	\begin{itemize}
		\item severity level: INFO
	\end{itemize}
\end{itemize}

\subsection{Context-Specific Valid Classes}

What types are valid in a specific context?\footnote{\emph{R-209-VALID-CLASSES}} 
Context can be an input stream, a data creation function, or an API.

\begin{itemize}
	\item \textbf{{\em DATA-CUBE-C-CONTEXT-SPECIFIC-VALID-CLASSES-01}}: For future versions of \emph{Data Cube}, out-dated classes can be marked as deprecated.
	\begin{itemize}
		\item severity level: INFO
	\end{itemize}
\end{itemize}

\begin{itemize}
	\item \textbf{{\em DCAT-C-CONTEXT-SPECIFIC-VALID-CLASSES-01}}: For future versions of \emph{DCAT}, out-dated classes can be marked as deprecated.
	\begin{itemize}
		\item severity level: INFO
	\end{itemize}
\end{itemize}

\begin{itemize}
	\item \textbf{{\em DISCO-C-CONTEXT-SPECIFIC-VALID-CLASSES-01}}: For future versions of \emph{Disco}, out-dated classes can be marked as deprecated.
	\begin{itemize}
		\item severity level: INFO
	\end{itemize}
\end{itemize}

\begin{itemize}
	\item \textbf{{\em PHDD-C-CONTEXT-SPECIFIC-VALID-CLASSES-01}}: For future versions of \emph{PHDD}, out-dated classes can be marked as deprecated.
	\begin{itemize}
		\item severity level: INFO
	\end{itemize}
\end{itemize}

\begin{itemize}
	\item \textbf{{\em SKOS-C-CONTEXT-SPECIFIC-VALID-CLASSES-01}}: For future versions of \emph{SKOS}, out-dated classes can be marked as deprecated.
	\begin{itemize}
		\item severity level: INFO
	\end{itemize}
\end{itemize}

\begin{itemize}
	\item \textbf{{\em XKOS-C-CONTEXT-SPECIFIC-VALID-CLASSES-01}}: For future versions of \emph{XKOS}, out-dated classes can be marked as deprecated.
	\begin{itemize}
		\item severity level: INFO
	\end{itemize}
\end{itemize}

\subsection{Context-Specific Valid Properties}

What properties can be used within this context?\footnote{\emph{R-210-VALID-PROPERTIES}} 
Context can be an data receipt function, data creation function, or API.

\begin{itemize}
	\item \textbf{{\em DATA-CUBE-C-CONTEXT-SPECIFIC-VALID-PROPERTIES-01}}: For future versions of \emph{Data Cube}, out-dated properties can be marked as deprecated.
	\begin{itemize}
		\item severity level: INFO
	\end{itemize}
\end{itemize}

\begin{itemize}
	\item \textbf{{\em DCAT-C-CONTEXT-SPECIFIC-VALID-PROPERTIES-01}}: For future versions of \emph{DCAT}, out-dated properties can be marked as deprecated.
	\begin{itemize}
		\item severity level: INFO
	\end{itemize}
\end{itemize}

\begin{itemize}
	\item \textbf{{\em DISCO-C-CONTEXT-SPECIFIC-VALID-PROPERTIES-01}}: For future versions of \emph{Disco}, out-dated properties can be marked as deprecated.
	\begin{itemize}
		\item severity level: INFO
	\end{itemize}
\end{itemize}

\begin{itemize}
	\item \textbf{{\em PHDD-C-CONTEXT-SPECIFIC-VALID-PROPERTIES-01}}: For future versions of \emph{PHDD}, out-dated properties can be marked as deprecated.
	\begin{itemize}
		\item severity level: INFO
	\end{itemize}
\end{itemize}

\begin{itemize}
	\item \textbf{{\em SKOS-C-CONTEXT-SPECIFIC-VALID-PROPERTIES-01}}: For future versions of \emph{SKOS}, out-dated properties can be marked as deprecated.
	\begin{itemize}
		\item severity level: INFO
	\end{itemize}
\end{itemize}

\begin{itemize}
	\item \textbf{{\em XKOS-C-CONTEXT-SPECIFIC-VALID-PROPERTIES-01}}: For future versions of \emph{XKOS}, out-dated properties can be marked as deprecated.
	\begin{itemize}
		\item severity level: INFO
	\end{itemize}
\end{itemize}

\subsection{Default Values}

\emph{Default values}\footnote{{\em R-31-DEFAULT-VALUES-OF-RDF-OBJECTS}, {\em R-38-DEFAULT-VALUES-OF-RDF-LITERALS}} for objects and literals are inferred automatically.
It should be possible to declare the default value for a given property, e.g. so that input forms can be pre-populated and to insert a required property that is missing in a web service call.

\begin{itemize}
	\item \textbf{{\em DISCO-C-DEFAULT-VALUES-01:}}
The value 'true' for the property {\em disco:isPublic} ({\em xsd:boolean}) indicates that the data set ({\em disco:LogicalDataSet}) can be accessed (usually downloaded) by anyone.
Per default, access to data sets should be restricted ('false').
\begin{itemize}
		\item severity level: INFO
	\end{itemize}
\end{itemize}

\subsection{Mathematical Operations}

Examples for {\em Mathematical Operations}\footnote{{\em R-42-MATHEMATICAL-OPERATIONS}, {\em R-41-STATISTICAL-COMPUTATIONS}} are the addition of two dates, the addition of days to a start date, and statistical computations (e.g. average, mean, sum).

\begin{itemize}
	\item \textbf{{\em DISCO-C-MATHEMATICAL-OPERATIONS-01:}}
  The sum of {\em disco:percentage} (datatype: {\em xsd:double}) values of all codes (represented as {\em skos:Concept}s) of a code list ({\em skos:ConceptScheme} or {\em skos:OrderedCollection}), serving as representation of a particular {\em disco:Variable}, must exactly be 100.
	\begin{itemize}
		\item severity level: ERROR
	\end{itemize}

	\item \textbf{{\em DISCO-C-MATHEMATICAL-OPERATIONS-02}}:
	For a given variable, the sum of the frequencies of all codes of the variable's code list has to be equal to the variable's total number of cases (summary statistics of the type 'number of cases').
	\begin{itemize}
		\item severity level: ERROR
	\end{itemize}
	
	\item \textbf{{\em DISCO-C-MATHEMATICAL-OPERATIONS-03}}:
	For a given variable, the sum of 'valid cases' and 'invalid cases' has to be equal to the total 'number of cases'.
	\begin{itemize}
		\item severity level: ERROR
	\end{itemize}
	
	\item \textbf{{\em DISCO-C-MATHEMATICAL-OPERATIONS-04}}:
	For a given variable, the total 'number of cases' value for the country 'All' must be equal to the sum of the total 'number of cases' value for each country.
	\begin{itemize}
		\item severity level: ERROR
	\end{itemize}
	
	\item \textbf{{\em DISCO-C-MATHEMATICAL-OPERATIONS-05}}:
	Minimum values do not have to be greater than maximum values (\emph{disco:SummaryStatistics}).
	\begin{itemize}
		\item severity level: ERROR
	\end{itemize}

\end{itemize}

\subsection{Language Tag Matching}

For particular data properties, values must be stated for predefined languages\footnote{{\em R-47-LANGUAGE-TAG-MATCHING}}.

\begin{itemize}
	\item \textbf{{\em DISCO-C-LANGUAGE-TAG-MATCHING-01:}}
There must be an English variable name ({\em skos:notation}) for each {\em disco:Variable} within {\em disco:LogicalDataSet}s.
\begin{itemize}
	\item severity level: INFO
\end{itemize}
\end{itemize}

\subsection{Language Tag Cardinality}

For particular data properties, values of predefined languages must be stated for determined number of times\footnote{{\em R-49-RDF-LITERALS-HAVING-AT-MOST-ONE-LANGUAGE-TAG}, {\em R-48-MISSING-LANGUAGE-TAGS}}.

\begin{itemize}

	\item \textbf{{\em DISCO-C-LANGUAGE-TAG-CARDINALITY-01:}}
  There must be at least one English {\em disco:questionText} for each {\em disco:Question} within {\em disco:LogicalDataSet}s.
\begin{itemize}
	\item severity level: INFO
\end{itemize}
  \item \textbf{{\em DISCO-C-LANGUAGE-TAG-CARDINALITY-02:}}
  There should be at most one English literal value for variable names ({\em skos:notation}, domain: {\em disco:Variable}).
\begin{itemize}
	\item severity level: INFO
\end{itemize}
  \item \textbf{{\em DISCO-C-LANGUAGE-TAG-CARDINALITY-03}}:
	For each question (\emph{disco:Question}), there must be at least one question text (\emph{disco:questionText}) associated with a language tag of an arbitrary language or with an English language tag.
\begin{itemize}
	\item severity level: INFO
\end{itemize}
\end{itemize}

\begin{itemize}

	\item \textbf{{\em SKOS-C-LANGUAGE-TAG-CARDINALITY-01\footnote{Corresponds to qSKOS Quality Issues - Labeling and Documentation Issues - Omitted or Invalid Language Tags}:}}
	Omitted or Invalid Language Tags:
Some controlled vocabularies contain literals in natural language, but without information what language has actually been used. Language tags might also not conform to language standards, such as RFC 3066. 
\begin{itemize}
	\item Implementation: Iteration over all triples in the vocabulary that have a predicate which is a (subclass of) \emph{rdfs:label} or \emph{skos:note}. 
  \item Severity level: WARNING
\end{itemize}

  \item \textbf{{\em SKOS-C-LANGUAGE-TAG-CARDINALITY-02\footnote{Corresponds to qSKOS Quality Issues - Labeling and Documentation Issues - Incomplete Language Coverage}:}}
	Incomplete Language Coverage:
	Some concepts in a thesaurus are labeled in only one language, some in multiple languages. It may be desirable to have each concept labeled in each of the languages that also are used on the other concepts. This is not always possible, but incompleteness of language coverage for some concepts can indicate shortcomings of the vocabulary. 
	\begin{itemize}
		\item Severity level: INFO
	\end{itemize}
	
	\item \textbf{{\em SKOS-C-LANGUAGE-TAG-CARDINALITY-03\footnote{Corresponds to qSKOS Quality Issues - Labeling and Documentation Issues - No Common Language}:}}
	No Common Language:
	Checks if all concepts have at least one common language, i.e. they have assigned at least one literal in the same language. 
	\begin{itemize}
		\item Severity level: INFO
	\end{itemize}
	
	\item \textbf{{\em SKOS-C-LANGUAGE-TAG-CARDINALITY-04\footnote{Corresponds to qSKOS Quality Issues - SKOS Semi-Formal Consistency Issues - Inconsistent Preferred Labels}:}}
	Inconsistent Preferred Labels:
  According to the SKOS reference document, "A resource has no more than one value of skos:prefLabel per language tag".  
	\begin{itemize}
	  \item Implementation:
		A SPARQL query is used to find concepts with at least two prefLabels. In a second step, the language tags of these prefLabels are analyzed and an ambiguity is detected if they are equal.  
		\item Severity level: INFO
	\end{itemize}
	
\end{itemize}

\subsection{Whitespace Handling}

Avoid whitespaces in literals neither leading nor trailing white spaces\footnote{\emph{R-50-WHITESPACE-HANDLING-OF-RDF-LITERALS}}.

\begin{itemize}
	\item \textbf{{\em DISCO-C-WHITESPACE-HANDLING-01:}} Delete whitespaces of series and study abstracts (\emph{dcterms:abstract}; domain: \emph{disco:StudyGroup}, \emph{disco:Study}) automatically.
	\begin{itemize}
		\item severity level: INFO
	\end{itemize}
\end{itemize}

\subsection{HTML Handling}

Check if all HTML tags, included in literals (of specific data properties within the context of specific classes)\footnote{\emph{R-51-HTML-HANDLING-OF-RDF-LITERALS}}, are closed properly.

\begin{itemize}
	\item \textbf{{\em DISCO-C-HTML-HANDLING-01:}} Check if all HTML tags, included in literals of all \emph{Disco} data properties, are closed properly.
		\begin{itemize}
		\item severity level: INFO
	\end{itemize}
	\item \textbf{{\em DISCO-C-HTML-HANDLING-02:}} Check if all HTML tags, included in literals of all data properties whose domains are \emph{Disco} classes, are closed properly.
		\begin{itemize}
		\item severity level: INFO
	\end{itemize}
\end{itemize}

\subsection{Conditional Properties}

If specific properties exist, then specific other properties must also be present\footnote{{\em R-71-CONDITIONAL-PROPERTIES}}.

\begin{itemize}
	\item \textbf{{\em DISCO-C-CONDITIONAL-PROPERTIES-01}}:
  If a {\em skos:Concept} represents a code (having a {\em skos:notation} property) and a category (having a {\em skos:prefLabel} property), 
then the property {\em disco:isValid} has to be stated indicating if the code is valid ('true') or missing ('false').
	\begin{itemize}
		\item severity level: ERROR
	\end{itemize}
	\item \textbf{{\em DISCO-C-CONDITIONAL-PROPERTIES-02}}:
	In order to get an overview over a series or a study either an abstract, a title, an alternative title, or links to external descriptions should be stated. 
	If the abstract (\emph{dcterms:abstract}) of a series (\emph{disco:StudyGroup}) and an external description of the series (\emph{disco:ddifile}) is missing, 
	a series title (\emph{dcterms:title}) or an alternative series title (\emph{dcterms:alternative}) has to be stated.
	\begin{itemize}
		\item severity level: WARNING
	\end{itemize}
	\item \textbf{{\em DISCO-C-CONDITIONAL-PROPERTIES-03}}:
	In order to get an overview over a series or a study either an abstract, a title, an alternative title, or links to external descriptions should be stated. 
	If the abstract (\emph{dcterms:abstract}) of a study (\emph{disco:Study}) and an external description of the study (\emph{disco:ddifile}) is missing, 
	a study title (\emph{dcterms:title}) or an alternative study title (\emph{dcterms:alternative}) has to be stated.
	\begin{itemize}
		\item severity level: WARNING
	\end{itemize}
	\item \textbf{{\em DISCO-C-CONDITIONAL-PROPERTIES-04}}:
	If the abstract (\emph{dcterms:abstract}) of a series (\emph{disco:StudyGroup}), an external description of the series (\emph{disco:ddifile}), 
	a series title (\emph{dcterms:title}), and an alternative series title (\emph{dcterms:alternative}) is missing, an error message should be shown.
	\begin{itemize}
		\item severity level: ERROR
	\end{itemize}
	\item \textbf{{\em DISCO-C-CONDITIONAL-PROPERTIES-05}}:
	If the abstract (\emph{dcterms:abstract}) of a study (\emph{disco:Study}), an external description of the study (\emph{disco:ddifile}), 
	a study title (\emph{dcterms:title}), and an alternative study title (\emph{dcterms:alternative}) is missing, an error message should be shown.
	\begin{itemize}
		\item severity level: ERROR
	\end{itemize}
	\item \textbf{{\em DISCO-C-CONDITIONAL-PROPERTIES-06}}:
	If a category statistics resource is connected with a code, it must be stated if the code is valid (\emph{disco:isValid}) and the code must be stated (\emph{skos:notation})
	\begin{itemize}
		\item severity level: ERROR
	\end{itemize}
\end{itemize}

\subsection{Recommended Properties}

Which properties are not necessarily required but recommended within a particular context\footnote{{\em R-72-RECOMMENDED-PROPERTIES}}.

\begin{itemize}
	\item \textbf{{\em DATA-CUBE-C-RECOMMENDED-PROPERTIES-01:}}
	\begin{itemize}
		\item severity level: INFO
	\end{itemize}
\end{itemize}

\begin{itemize}
	\item \textbf{{\em DCAT-C-RECOMMENDED-PROPERTIES-01:}}
	\begin{itemize}
		\item severity level: INFO
	\end{itemize}
\end{itemize}

\begin{itemize}
	\item \textbf{{\em DISCO-C-RECOMMENDED-PROPERTIES-01:}}
  The property {\em skos:notation} is not mandatory for {\em disco:Variable}s, but recommended to indicate variable names.
	\begin{itemize}
		\item severity level: INFO
	\end{itemize}
\end{itemize}

\begin{itemize}
	\item \textbf{{\em PHDD-C-RECOMMENDED-PROPERTIES-01:}}
	\begin{itemize}
		\item severity level: INFO
	\end{itemize}
\end{itemize}

\begin{itemize}
	\item \textbf{{\em SKOS-C-RECOMMENDED-PROPERTIES-01:}}
	\begin{itemize}
		\item severity level: INFO
	\end{itemize}
\end{itemize}

\begin{itemize}
	\item \textbf{{\em XKOS-C-RECOMMENDED-PROPERTIES-01:}}
	\begin{itemize}
		\item severity level: INFO
	\end{itemize}
\end{itemize}

\subsection{Handle RDF Collections}

Examples of the \emph{Handle RDF Collections}\footnote{\emph{R-120-HANDLE-RDF-COLLECTIONS}} constraint are: a collection must have a specific size; the first/last element of a given list must be a specific literal; the elements of collections are compared; are collections identical?; actions on RDF lists\footnote{See \url{http://www.snee.com/bobdc.blog/2014/04/rdf-lists-and-sparql.html}}; the 2. list element must be equal to 'XXX'; does the list have more than 10 elements?

\begin{itemize}
	\item \textbf{{\em DISCO-C-HANDLE-RDF-COLLECTIONS-01}}: Have comparable \emph{disco:Variable}s the same number of codes in their code lists?
	\begin{itemize}
		\item severity level: INFO
	\end{itemize}
	\item \textbf{{\em DISCO-C-HANDLE-RDF-COLLECTIONS-02}}: Does the actual number of \emph{disco:Variable}s within an (un)ordered collection of a given \emph{disco:LogicalDataSet} match the expected number? 
	\begin{itemize}
		\item severity level: INFO
	\end{itemize}
\end{itemize}

\subsection{Value is Valid for Datatype}

Make sure that a value is valid for its datatype.
It has to be ensured, e.g., that a date is really a date, or that a \emph{xsd:nonNegativeInteger} value is not negative. 

\begin{itemize}
	\item \textbf{{\em DISCO-C-VALUE-IS-VALID-FOR-DATATYPE-01}}: 
	Check if all literal values of properties used within the \emph{Disco} context of the datatype {\em xsd:date} (e.g. {\em disco:startDate}, {\em disco:endDate}, {\em dcterms:date}) are really of the datatype {\em xsd:date}.
	\begin{itemize}
		\item severity level: ERROR
	\end{itemize}
	\item \textbf{{\em DISCO-C-VALUE-IS-VALID-FOR-DATATYPE-02}}: 
	Frequencies (disco:frequency) cannot be negative, i.e., must correspond  to the XML Schema datatype \emph{xsd:nonNegativeInteger}.
	\begin{itemize}
		\item severity level: ERROR
	\end{itemize}
\end{itemize}

\begin{itemize}
	\item \textbf{{\em DATA-CUBE-C-VALUE-IS-VALID-FOR-DATATYPE-01:}}
	Datatype consistency (\emph{IC-0} \cite{CyganiakReynolds2014}) -  
	The RDF graph must be consistent under RDF D-entailment using a datatype map containing all the datatypes used within the graph. 
	\begin{itemize}
		\item severity level: ERROR
	\end{itemize}
\end{itemize}

\subsection[Use Sub-Super Relations in Validation] {Use Sub-Super Relations in Validation\footnote{\emph{R-224-USE-SUB-SUPER-RELATIONS-IN-VALIDATION}}}


The validation of instances data (direct or indirect) exploits the sub-class or sub-property link in a given ontology.
This validation can indicate when the data is verbose (redundant) or expressed at a too general level, and could be improved.
If \emph{dcterms:date} and one of its sub-properties \emph{dcterms:created} or \emph{dcterms:issued} are present, e.g., check that the value in \emph{dcterms:date} is not redundant with \emph{dcterms:created} or \emph{dcterms:issued} for ingestion.

\begin{itemize}
	\item \textbf{{\em DISCO-C-USE-SUB-SUPER-RELATIONS-IN-VALIDATION-01}}: 
	If one or more \emph{dcterms:coverage} properties are present, suggest the use of one of its sub-properties \emph{dcterms:spatial} or \emph{dcterms:temporal}.
		\begin{itemize}
		\item severity level: INFO
	\end{itemize}
	\item \textbf{{\em DISCO-C-USE-SUB-SUPER-RELATIONS-IN-VALIDATION-02}}: 
	If the \emph{dcterms:contributor} property is present, suggest the use of one of its sub-properties, e.g. \emph{disco:fundedBy}.
		\begin{itemize}
		\item severity level: INFO
	\end{itemize}
\end{itemize}

\subsection{Cardinality Shortcuts}

In most library applications, cardinality shortcuts tend to appear in pairs, with repeatable/non-repeatable establishing maximum cardinality and optional/mandatory establishing minimum cardinality.
These are shortcuts for more detailed \emph{cardinality restrictions}.

\subsection{Aggregations}

Some constraints require aggregating multiple values, especially via \emph{COUNT}, \emph{MIN} and \emph{MAX}.

\begin{itemize}
	\item \textbf{{\em DISCO-C-AGGREGATION-01}}: calculate the number of theoretical concepts in the thematic classification of a given study.
	\begin{itemize}
		\item severity level: INFO
	\end{itemize}
  \item \textbf{{\em DISCO-C-AGGREGATION-02}}: calculate the number of variables of a data set.
	\begin{itemize}
		\item severity level: INFO
	\end{itemize}
	\item \textbf{{\em DISCO-C-AGGREGATION-03}}: calculate the number of questions in a given questionnaire.
	\begin{itemize}
		\item severity level: INFO
	\end{itemize}
	\item \textbf{{\em DISCO-C-AGGREGATION-04}}: the number of codes of a given variable must be below a maximum value.
	\begin{itemize}
		\item severity level: INFO
	\end{itemize}
	\item \textbf{{\em DISCO-C-AGGREGATION-05}}: the number of questions of a given questionnaire must exactly be a given value.
	\begin{itemize}
		\item severity level: INFO
	\end{itemize}
	\item \textbf{{\em DISCO-C-AGGREGATION-06}}: the sum of percentages of all codes of a given variable must be 100.
	\begin{itemize}
		\item severity level: INFO
	\end{itemize}
  \item \textbf{{\em DISCO-C-AGGREGATION-07}}: the absolute frequency of all valid codes of a given variable must be equal to a given value.
	\begin{itemize}
		\item severity level: INFO
	\end{itemize}
\end{itemize} 

\subsection{Provenance}

\begin{itemize}
	\item \textbf{{\em DISCO-C-PROVENANCE-01}}: 
	Series should have provenance information (\emph{dcterms:provenance}).
	\begin{itemize}
		\item severity level: INFO
	\end{itemize}
	\item \textbf{{\em DISCO-C-PROVENANCE-02}}: 
	Studies should have provenance information (\emph{dcterms:provenance}).
	\begin{itemize}
		\item severity level: INFO
	\end{itemize}
	\item \textbf{{\em DISCO-C-PROVENANCE-03}}: 
	Data sets should have provenance information (\emph{dcterms:provenance}).
	\begin{itemize}
		\item severity level: INFO
	\end{itemize}
	\item \textbf{{\em DISCO-C-PROVENANCE-04}}: 
	Data files should have provenance information (\emph{dcterms:provenance}).
	\begin{itemize}
		\item severity level: INFO
	\end{itemize}
\end{itemize} 
 
\subsection{Comparison}
\setcounter{subsection}{76}

\begin{itemize}
	\item \textbf{{\em DISCO-C-COMPARISON-VARIABLES-01}}: 
	are compared variables represented in a compatible way, i.e. are the variables' code lists theoretically comparable?
		\begin{itemize}
		\item severity level: WARNING
	\end{itemize}
	\item \textbf{{\em DISCO-C-COMPARISON-VARIABLES-02}}: 
	are variable definitions (\emph{dcterms:description}) available for each variable (\emph{disco:Variable}) to compare?
		\begin{itemize}
		\item severity level: ERROR
	\end{itemize}
	\item \textbf{{\em DISCO-C-COMPARISON-VARIABLES-03}}:
	are code lists structured properly for each variable (\emph{disco:Variable}) to compare?
		\begin{itemize}
		\item severity level: ERROR
	\end{itemize}
	\item \textbf{{\em DISCO-C-COMPARISON-VARIABLES-04}}:
	is for each code (for each variable (\emph{disco:Variable}) to compare) an associated category (a human-readable label) specified?
		\begin{itemize}
		\item severity level: INFO
	\end{itemize}
	\item \textbf{{\em DISCO-C-COMPARISON-VARIABLES-05}}:
	each (\emph{disco:Variable}) to compare must be present.
	\begin{itemize}
		\item severity level: ERROR
	\end{itemize}
\end{itemize}

\subsection{Data Model Consistency}
Is the data consistent with the intended semantics of the data model?
Such validation rules ensure the integrity of the data according to the data model.

\begin{itemize}
	\item \textbf{{\em DISCO-C-DATA-MODEL-CONSISTENCY-01}}: 
	Codes (\emph{skos:Concept}) are ordered and therefore have fixed positions in an ordered collection (\emph{skos:OrderedCollection}) within a variable representation.
	The cumulative percentage of the current code is the cumulative percentage of the previous code (\emph{disco:cumulativePercentage})
	plus the percentage value (\emph{disco:percentage}) of the current code. 
		\begin{itemize}
		\item severity level: ERROR
	\end{itemize}
	\item \textbf{{\em DISCO-C-DATA-MODEL-CONSISTENCY-02}}: 
	The cumulative percentage (\emph{disco:cumulativePercentage}) of the last code must be 100. 
		\begin{itemize}
		\item severity level: ERROR
	\end{itemize}
	\item \textbf{{\em DISCO-C-DATA-MODEL-CONSISTENCY-03}}: 
	The number of valid cases (\emph{disco:SummaryStatistics} of the type (\emph{disco:summaryStatisticType}) \emph{ddicv-sumstats:ValidCases}) 
	for a particular variable must exactly be the sum of all frequencies of all valid cases (\emph{disco:inValid} of \emph{skos:Concept} is true).
		\begin{itemize}
		\item severity level: ERROR
	\end{itemize}
	\item \textbf{{\em DISCO-C-DATA-MODEL-CONSISTENCY-04}}: 
	The number of invalid cases (\emph{disco:SummaryStatistics} of the type (\emph{disco:summaryStatisticType}) \emph{ddicv-sumstats:InvalidCases})
	for a particular variable must exactly be the sum of all frequencies of all invalid cases (\emph{disco:inValid} of \emph{skos:Concept} is false).
		\begin{itemize}
		\item severity level: ERROR
	\end{itemize}
	\item \textbf{{\em DISCO-C-DATA-MODEL-CONSISTENCY-05}}: 
	The total number of cases (\emph{rdf:value} of the \emph{disco:SummaryStatistics} resource of the type (\emph{disco:summaryStatisticType}) \emph{ddicv-sumstats:NumberOfCases}) 
	for a particular variable must exactly be the number of valid cases plus the number of invalid cases.
		\begin{itemize}
		\item severity level: ERROR
	\end{itemize}
	\item \textbf{{\em DISCO-C-DATA-MODEL-CONSISTENCY-06}}: 
	Some summary statistics types can only be calculated for given variable types. 
	It is not possible to compute minimum values for string variables.
		\begin{itemize}
		\item severity level: ERROR
	\end{itemize}
	\item \textbf{{\em DISCO-C-DATA-MODEL-CONSISTENCY-07}}: 
	Some summary statistics types can only be calculated for given variable types. 
	It is not possible to compute mean values for categorical variables, only for metric variables.
		\begin{itemize}
		\item severity level: ERROR
	\end{itemize}
\end{itemize}

\begin{itemize}
	\item \textbf{{\em DATA-CUBE-C-DATA-MODEL-CONSISTENCY-01:}}
	Only attributes may be optional (\emph{IC-6} \cite{CyganiakReynolds2014}) - 
	The only components of a \emph{qb:DataStructureDefinition} that may be marked as optional, using \emph{qb:componentRequired} are attributes. 
	\begin{itemize}
		\item severity level: WARNING
	\end{itemize}
	\item \textbf{{\em DATA-CUBE-C-DATA-MODEL-CONSISTENCY-02:}}
	Slice Keys consistent with DSD (\emph{IC-8} \cite{CyganiakReynolds2014}) -
	Every \emph{qb:componentProperty} on a \emph{qb:SliceKey} must also be declared as a \emph{qb:component} of the associated \emph{qb:DataStructureDefinition}. 
	\begin{itemize}
		\item severity level: ERROR
	\end{itemize}
	\item \textbf{{\em DATA-CUBE-C-DATA-MODEL-CONSISTENCY-03:}} 
	Slice dimensions complete (\emph{IC-10} \cite{CyganiakReynolds2014}) - 
	Every \emph{qb:Slice} must have a value for every dimension declared in its \emph{qb:sliceStructure}. 
	\begin{itemize}
		\item severity level: ERROR
	\end{itemize}
		\item \textbf{{\em DATA-CUBE-C-DATA-MODEL-CONSISTENCY-04:}}
	All dimensions required (\emph{IC-11} \cite{CyganiakReynolds2014}) - 
	Every \emph{qb:Observation} has a value for each dimension declared in its associated \emph{qb:DataStructureDefinition}. 
	\begin{itemize}
		\item severity level: ERROR
	\end{itemize}
		\item \textbf{{\em DATA-CUBE-C-DATA-MODEL-CONSISTENCY-05:}}
	No duplicate observations (\emph{IC-12} \cite{CyganiakReynolds2014}) - 
	No two \emph{qb:Observations} in the same \emph{qb:DataSet} may have the same value for all dimensions. 
	\begin{itemize}
		\item severity level: WARNING
	\end{itemize}
	\item \textbf{{\em DATA-CUBE-C-DATA-MODEL-CONSISTENCY-06:}}
	Required attributes (\emph{IC-13} \cite{CyganiakReynolds2014}) -
	Every \emph{qb:Observation} has a value for each declared attribute that is marked as required. 
	\begin{itemize}
		\item severity level: ERROR
	\end{itemize}
	\item \textbf{{\em DATA-CUBE-C-DATA-MODEL-CONSISTENCY-07:}}
	All measures present (\emph{IC-14} \cite{CyganiakReynolds2014}) -
	In a \emph{qb:DataSet} which does not use a Measure dimension then each individual \emph{qb:Observation} must have a value for every declared measure. 
	\begin{itemize}
		\item severity level: ERROR
	\end{itemize}
	\item \textbf{{\em DATA-CUBE-C-DATA-MODEL-CONSISTENCY-08:}}
	Measure dimension consistent (\emph{IC-15} \cite{CyganiakReynolds2014}) -
	In a \emph{qb:DataSet} which uses a Measure dimension then each qb:Observation must have a value for the measure corresponding to its given \emph{qb:measureType}. 
	\begin{itemize}
		\item severity level: ERROR
	\end{itemize}
	\item \textbf{{\em DATA-CUBE-C-DATA-MODEL-CONSISTENCY-09:}}
	Single measure on measure dimension observation (\emph{IC-16} \cite{CyganiakReynolds2014}) -
	In a \emph{qb:DataSet} which uses a Measure dimension then each \emph{qb:Observation} must only have a value for one measure (by \emph{IC-15} this will be the measure corresponding to its \emph{qb:measureType}). 
	\begin{itemize}
		\item severity level: ERROR
	\end{itemize}
	\item \textbf{{\em DATA-CUBE-C-DATA-MODEL-CONSISTENCY-10:}}
	All measures present in measures dimension cube (\emph{IC-17} \cite{CyganiakReynolds2014}) -
	In a \emph{qb:DataSet} which uses a Measure dimension then if there is a Observation for some combination of non-measure dimensions then there must be other Observations with the same non-measure dimension values for each of the declared measures. 
	\begin{itemize}
		\item severity level: ERROR
	\end{itemize}
	\item \textbf{{\em DATA-CUBE-C-DATA-MODEL-CONSISTENCY-11:}}
	Consistent data set links (\emph{IC-18} \cite{CyganiakReynolds2014}) -
	If a \emph{qb:DataSet} \emph{D} has a \emph{qb:slice} \emph{S}, and \emph{S} has an \emph{qb:observation} \emph{O}, then the \emph{qb:dataSet} corresponding to \emph{O} must be \emph{D}. 
	\begin{itemize}
		\item severity level: WARNING
	\end{itemize}
\end{itemize}

\begin{itemize}

	\item \textbf{{\em SKOS-C-DATA-MODEL-CONSISTENCY-01\footnote{Corresponds to qSKOS Quality Issues - SKOS Semi-Formal Consistency Issues - Relation Clashes}:}}
	Relation Clashes:
  Covers condition S27 from the SKOS reference document, that has not been defined formally. 
	\begin{itemize}
	  \item Implementation:
		In a first step, all pairs of concepts are found that are associatively connected, using a SPARQL query. In the second step, a graph is created, containing only hierarchically related concepts and the respective relations. For each concept pair from the first step, we check for a path in the graph from step two. If such a path is found, a clash has been identified and the causing concepts are returned. 
		\item Severity level: INFO
	\end{itemize}
	
	\item \textbf{{\em SKOS-C-DATA-MODEL-CONSISTENCY-02\footnote{Corresponds to qSKOS Quality Issues - SKOS Semi-Formal Consistency Issues - Mapping Clashes}:}}
	Mapping Clashes:
  Covers condition S46 from the SKOS reference document, that has not been defined formally. 
	\begin{itemize}
	  \item Implementation:
		Can be solved by issuing a SPARQL query. 
		\item Severity level: INFO
	\end{itemize}
	
	\item \textbf{{\em SKOS-C-DATA-MODEL-CONSISTENCY-03\footnote{Corresponds to qSKOS Quality Issues - SKOS Semi-Formal Consistency Issues - Mapping Relations Misuse}:}}
	Mapping Relations Misuse:
  According to the SKOS reference documentation, mapping relations (e.g., \emph{skos:broadMatch} or \emph{skos:relatedMatch}) should be asserted to concepts being members of different concept schemes. This check finds concepts that are related by a mapping property and are either members of the same concept scheme or members of no concept scheme at all. 
	\begin{itemize}
		\item Severity level: INFO
	\end{itemize}
	
\end{itemize}

\subsection{Structure}

SKOS is based on RDF, which is a graph-based data model. Therefore we can concentrate on the vocabulary's graph-based structure for assessing the quality of SKOS vocabularies and apply graph- and network-analysis techniques. 

\begin{itemize}
	\item \textbf{{\em DISCO-C-STRUCTURE-01}}: there must be exactly one root in the hierarchy of DDI concepts. 
	\begin{itemize}
		\item severity level: ERROR
	\end{itemize}
\end{itemize} 

\begin{itemize}
	\item \textbf{{\em DATA-CUBE-C-STRUCTURE-01:}}
	Codes from hierarchy (\emph{IC-20} \cite{CyganiakReynolds2014}) -  
	If a dimension property has a \emph{qb:HierarchicalCodeList} with a non-blank \emph{qb:parentChildProperty} then the value of that dimension property on every \emph{qb:Observation} must be reachable from a root of the hierarchy using zero or more hops along the \emph{qb:parentChildProperty} links. 
	\begin{itemize}
		\item severity level: ERROR
	\end{itemize}
	\item \textbf{{\em DATA-CUBE-C-STRUCTURE-02:}}
	Codes from hierarchy (inverse) (\emph{IC-21} \cite{CyganiakReynolds2014}) -  
	If a dimension property has a \emph{qb:HierarchicalCodeList} with an inverse \emph{qb:parentChildProperty} then the value of that dimension property on every \emph{qb:Observation} must be reachable from a root of the hierarchy using zero or more hops along the inverse \emph{qb:parentChildProperty} links. 
	\begin{itemize}
		\item severity level: ERROR
	\end{itemize}
\end{itemize}

\begin{itemize}
	\item \textbf{{\em SKOS-C-STRUCTURE-01\footnote{Corresponds to qSKOS Quality Issues - Structural Issues - Orphan Concepts}:}}
	Orphan Concepts:
  An orphan concept is a concept without any associative or hierarchical relations. It might have attached literals like e.g., labels, but is not connected to any other resource, lacking valuable context information. A controlled vocabulary that contains many orphan concepts is less usable for search and retrieval use cases, because, e.g., no hierarchical query expansion can be performed on search terms to find documents with more general content. 
	\begin{itemize}
		\item Implementation: Iteration over all concepts in the vocabulary and returning that don't have associated resources using (sub-properties of) \emph{skos:semanticRelation}. 
		\item Severity level: WARNING
\end{itemize}
	\item \textbf{{\em SKOS-C-STRUCTURE-02\footnote{Corresponds to qSKOS Quality Issues - Structural Issues - Disconnected Concept Clusters}:}}
	Disconnected Concept Clusters:
	Checking the connectivity of the graph, it is possible to identify all weakly connected components. These datasets form "islands" in the vocabulary and might be caused by incomplete data acquisition, "forgotten" test data, outdated terms and the like. 
	\begin{itemize}
		\item Implementation: Creation of an undirected graph that includes all non-orphan concepts as nodes and all semantic relations as edges. Tarjan's algorithm then finds and returns all weakly connected components.
		\item Severity level: INFO
\end{itemize}
	\item \textbf{{\em SKOS-C-STRUCTURE-03\footnote{Corresponds to qSKOS Quality Issues - Structural Issues - Cyclic Hierarchical Relations}:}}
	Cyclic Hierarchical Relations: 
	Although perfectly consistent with the SKOS data model, cyclic relations may reveal a logical problem in the thesaurus. Consider the following example: "decision" $\rightarrow$ "problem resolution" $\rightarrow$ "problem" ($\rightarrow$ "decision": here the cycle is closed). The concepts are connected using \emph{skos:broader} relationships (indicated with "$\rightarrow$"). Due to the fact that a thesaurus is in many cases a product of consensus between the contributors (or just the decision of one dedicated thesaurus manager), it will be almost impossible to automatically resolve the cycle (i.e. deleting an edge). 
	\begin{itemize}
		\item Implementation: Construction of a graph having all concepts as nodes and the set of edges being \emph{skos:broader} relations. 
		\item Severity level: WARNING
\end{itemize}
	\item \textbf{{\em SKOS-C-STRUCTURE-04\footnote{Corresponds to qSKOS Quality Issues - Structural Issues - Valueless Associative Relations}:}}
	Valueless Associative Relations:
	Two concepts are sibling, but also connected by an associative relation. In that context, the associative relation is not necessary. See ISO\_DIS\_25964-1, 11.3.2.2 
	\begin{itemize}
		\item Implementation:
		Identification of all pairs of concepts that have the same broader or narrower concepts, i.e. they are "sibling terms". All siblings that are related by a \emph{skos:related} property are returned. 
	  \item Severity level: INFO
\end{itemize}
	\item \textbf{{\em SKOS-C-STRUCTURE-05\footnote{Corresponds to qSKOS Quality Issues - Structural Issues - Solely Transitively Related Concepts}:}}
	Solely Transitively Related Concepts:
	\emph{skos:broaderTransitive} and \emph{skos:narrowerTransitive} are, according to the SKOS reference document, "not used to make assertions", so they should not be the only relations hierarchically relating two concepts. 
	\begin{itemize}
		\item Implementation: Identification of all concept pairs that are related by \emph{skos:broaderTransitive} or \emph{skos:narrowerTransitive} properties but not by their \emph{skos:broader} and \emph{skos:narrower} subproperties. 
		\item Severity level: INFO
\end{itemize}
	\item \textbf{{\em SKOS-C-STRUCTURE-06\footnote{Corresponds to qSKOS Quality Issues - Structural Issues - Unidirectionally Related Concepts}:}}
	Unidirectionally Related Concepts:
	Reciprocal relations (e.g., \emph{broader/narrower, related, hasTopConcept/topConceptOf}) should be included in the controlled vocabularies to achieve better search results using SPARQL in systems without reasoner support. 
	\begin{itemize}
		\item Implementation:
		This issue is checked without inference of \emph{owl:inverseOf} properties. We iterate over all triples and check for each property if an inverse property is defined in the SKOS ontology and if the respective statement using this property is included in the vocabulary. If not, the resources associated with this property are returned. 
	  \item Severity level: INFO
\end{itemize}
	\item \textbf{{\em SKOS-C-STRUCTURE-07\footnote{Corresponds to qSKOS Quality Issues - Structural Issues - Omitted Top Concepts}:}}
	Omitted Top Concepts:
	A vocabulary should provide "entry points" to the data to provide “efficient access” (SKOS primer) and guidance for human users. 
	\begin{itemize}
		\item Implementation:
		For every ConceptScheme in the controlled vocabulary, a SPARQL query is issued finding resources that are associated with this ConceptScheme by one of the properties \emph{skos:hasTopConcept} or \emph{skos:topConceptOf}. Top concepts are also concepts having no broader concept. 
	  \item Severity level: WARNING
\end{itemize}
	\item \textbf{{\em SKOS-C-STRUCTURE-08\footnote{Corresponds to qSKOS Quality Issues - Structural Issues - Top Concepts Having Broader Concepts}:}}
	Top Concepts Having Broader Concepts:
	Concepts "internal to the tree" should not be indicated as top concepts.
	\begin{itemize}
		\item Implementation:
		A SPARQL query finds all top concepts (being defined by one of the properties \emph{skos:hasTopConcept} or \emph{skos:topConceptOf}) having associated a broader concept. 
	  \item Severity level: ERROR
\end{itemize}
	\item \textbf{{\em SKOS-C-STRUCTURE-09\footnote{Corresponds to qSKOS Quality Issues - Structural Issues - Hierarchical Redundancy}:}}
	Hierarchical Redundancy:
	As stated in the SKOS reference document, \emph{skos:broader} and \emph{skos:narrower} are not transitive properties. However, they are sub-properties of \emph{skos:broaderTransitive} and \emph{skos:narrowerTransitive} which enables inference of a "transitive closure". This, in fact, leaves it up to the user to interpret whether a vocabulary's hierarchical structure is seen as transitive or not. In the former case, this check can be useful. It finds pairs of concepts (A,B) that are directly hierarchically related but there also exits an hierarchical path through a concept C that connects A and B. 
	\begin{itemize}
		\item Severity level: INFO
	\end{itemize}
	\item \textbf{{\em SKOS-C-STRUCTURE-10\footnote{Corresponds to qSKOS Quality Issues - Structural Issues - Reflexive Relations}:}}
	Reflexive Relations:
	Concepts related to themselves. 
	\begin{itemize}
		\item Severity level: WARNING
	\end{itemize}
\end{itemize}

\subsection{Labeling and Documentation}

\begin{itemize}
	\item \textbf{{\em DISCO-C-LABELING-AND-DOCUMENTATION-01}}: 
	Series should be described (\emph{dcterms:description}).
	\begin{itemize}
		\item severity level: INFO
	\end{itemize}
	\item \textbf{{\em DISCO-C-LABELING-AND-DOCUMENTATION-02}}: 
	Studies should be described (\emph{dcterms:description}).
	\begin{itemize}
		\item severity level: INFO
	\end{itemize}
	\item \textbf{{\em DISCO-C-LABELING-AND-DOCUMENTATION-03}}: 
	Data sets should be described (\emph{dcterms:description}).
	\begin{itemize}
		\item severity level: INFO
	\end{itemize}
	\item \textbf{{\em DISCO-C-LABELING-AND-DOCUMENTATION-04}}: 
	Data files should be described (\emph{dcterms:description}).
	\begin{itemize}
		\item severity level: INFO
	\end{itemize}
	\item \textbf{{\em DISCO-C-LABELING-AND-DOCUMENTATION-05}}: 
	Instruments should be described (\emph{dcterms:description}).
	\begin{itemize}
		\item severity level: INFO
	\end{itemize}
	\item \textbf{{\em DISCO-C-LABELING-AND-DOCUMENTATION-06}}: 
	Variables should be described (\emph{dcterms:description}).
	\begin{itemize}
		\item severity level: INFO
	\end{itemize}
\end{itemize}

\begin{itemize}
	\item \textbf{{\em SKOS-C-LABELING-AND-DOCUMENTATION-01\footnote{Corresponds to qSKOS Quality Issues - Labeling and Documentation Issues - Undocumented Concepts}:}}
	Undocumented Concepts:
  The SKOS standard defines a number of properties useful for documenting the meaning of the concepts in a thesaurus also in a human-readable form. Intense use of these properties leads to a well-documented thesaurus which should also improve its quality.  
	\begin{itemize}
	  \item Implementation:
		Iteration over all concepts in the vocabulary and find those not using one of \emph{skos:note}, \emph{skos:changeNote}, \emph{skos:definition}, \emph{skos:editorialNote}, \emph{skos:example}, \emph{skos:historyNote}, or \emph{skos:scopeNote}.
		\item Severity level: INFO
	\end{itemize}
	\item \textbf{{\em SKOS-C-LABELING-AND-DOCUMENTATION-02\footnote{Corresponds to qSKOS Quality Issues - Labeling and Documentation Issues - Overlapping Labels}:}}
	Overlapping Labels:
	This is a generalization of a recommendation in the SKOS primer, that “no two concepts have the same preferred lexical label in a given language when they belong to the same concept scheme”. This could indicate missing disambiguation information and thus lead to problems in autocompletion application. 
	\begin{itemize}
		\item Severity level: INFO
	\end{itemize}
	\item \textbf{{\em SKOS-C-LABELING-AND-DOCUMENTATION-03\footnote{Corresponds to qSKOS Quality Issues - Labeling and Documentation Issues - Missing Labels}:}}
	Missing Labels:
	To make the vocabulary more convenient for humans to use, instances of SKOS classes (Concept, ConceptScheme, Collection) should be labeled using e.g., \emph{skos:prefLabel}, \emph{altLabel}, \emph{rdfs:label}, \emph{dc:title}. 
	\begin{itemize}
		\item Severity level: INFO
	\end{itemize}
	\item \textbf{{\em SKOS-C-LABELING-AND-DOCUMENTATION-04\footnote{Corresponds to qSKOS Quality Issues - Labeling and Documentation Issues - Unprintable Characters in Labels}:}}
	Unprintable Characters in Labels:
	\emph{pref/alt/hiddenlabels} contain characters that are not alphanumeric characters or blanks.
	\begin{itemize}
		\item Severity level: INFO
	\end{itemize}
	\item \textbf{{\em SKOS-C-LABELING-AND-DOCUMENTATION-05\footnote{Corresponds to qSKOS Quality Issues - Labeling and Documentation Issues - Empty Labels}:}}
	Empty Labels:
	Labels also need to contain textual information to be useful, thus we find all SKOS labels with length 0 (after removing whitespaces). 
	\begin{itemize}
		\item Severity level: INFO
	\end{itemize}
	\item \textbf{{\em SKOS-C-LABELING-AND-DOCUMENTATION-06\footnote{Corresponds to qSKOS Quality Issues - Labeling and Documentation Issues - Ambiguous Notation References}:}}
	Ambiguous Notation References:
	Concepts within the same concept scheme should not have identical \emph{skos:notation} literals. 
	\begin{itemize}
		\item Severity level: INFO
	\end{itemize}
\end{itemize}

\subsection{Vocabulary}
Vocabularies should not invent any new terms or use deprecated elements. 

\begin{itemize}
	\item \textbf{\em DATA-CUBE-C-VOCABULARY-01} 
	\begin{itemize}
		\item Severity level: ERROR
	\end{itemize}
\end{itemize}

\begin{itemize}
	\item \textbf{\em DCAT-C-VOCABULARY-01} 
	\begin{itemize}
		\item Severity level: ERROR
	\end{itemize}
\end{itemize}

\begin{itemize}
	\item \textbf{\em DISCO-C-VOCABULARY-01} 
	\begin{itemize}
		\item Severity level: ERROR
	\end{itemize}
\end{itemize}

\begin{itemize}
	\item \textbf{\em PHDD-C-VOCABULARY-01} 
	\begin{itemize}
		\item Severity level: ERROR
	\end{itemize}
\end{itemize}

\begin{itemize}
	\item \textbf{{\em SKOS-C-VOCABULARY-01\footnote{Corresponds to qSKOS Quality Issues - Linked Data Specific Issues - Undefined SKOS Resources}:}}
	Undefined SKOS Resources:
	The vocabulary should not invent any new terms within the SKOS namespace or use deprecated SKOS elements. 
	\begin{itemize}
		\item Severity level: ERROR
	\end{itemize}
\end{itemize}

\begin{itemize}
	\item \textbf{\em XKOS-C-VOCABULARY-01} 
	\begin{itemize}
		\item Severity level: ERROR
	\end{itemize}
\end{itemize}

\subsection{HTTP URI Scheme Violation}

\begin{itemize}
	\item \textbf{\em DISCO-C-HTTP-URI-SCHEME-VIOLATION\footnote{Corresponds to qSKOS Quality Issues - Linked Data Specific Issues - HTTP URI Scheme Violation}}: 
	In the context of Linked Data, we restrict ourselves to using HTTP URIs only and avoid other URI schemes such as URNs and DOIs.
	\begin{itemize}
		\item Severity level: ERROR
	\end{itemize}
\end{itemize}

\begin{itemize}
	\item \textbf{\em DATA-CUBE-C-HTTP-URI-SCHEME-VIOLATION\footnote{Corresponds to qSKOS Quality Issues - Linked Data Specific Issues - HTTP URI Scheme Violation}}: 
	In the context of Linked Data, we restrict ourselves to using HTTP URIs only and avoid other URI schemes such as URNs and DOIs.
	\begin{itemize}
		\item Severity level: ERROR
	\end{itemize}
\end{itemize}

\begin{itemize}
	\item \textbf{\em PHDD-C-HTTP-URI-SCHEME-VIOLATION\footnote{Corresponds to qSKOS Quality Issues - Linked Data Specific Issues - HTTP URI Scheme Violation}}: 
	In the context of Linked Data, we restrict ourselves to using HTTP URIs only and avoid other URI schemes such as URNs and DOIs.
	\begin{itemize}
		\item Severity level: ERROR
	\end{itemize}
\end{itemize}

\begin{itemize}
	\item \textbf{\em SKOS-C-HTTP-URI-SCHEME-VIOLATION\footnote{Corresponds to qSKOS Quality Issues - Linked Data Specific Issues - HTTP URI Scheme Violation}}: 
	In the context of Linked Data, we restrict ourselves to using HTTP URIs only and avoid other URI schemes such as URNs and DOIs.
	\begin{itemize}
		\item Severity level: ERROR
	\end{itemize}
\end{itemize}

\section{Conclusion}

We identified and published by today 81 types of constraints  that are required by various stakeholders for data applications.
In close collaboration with several domain experts for the social, behavioral, and economic sciences (SBE), we formulated constraints on common SBE vocabularies and classified them according to their severity level.

\bibliography{../../../literature/literature}{}
\bibliographystyle{plain}
\setcounter{tocdepth}{1}
\end{document}